\documentclass[aps,twocolumn,preprintnumbers,amsmath,amssymb,floatfix,prb,superscriptaddress]{revtex4}
\usepackage{graphicx}
\usepackage{rotating}
\usepackage{comment}
\usepackage{dcolumn}
\usepackage{makeidx}
\usepackage{color}
\usepackage{amsmath}
\usepackage{amsfonts}
\usepackage{braket}
\usepackage{bm}
\usepackage{float}
\usepackage[normalem]{ulem}

\usepackage{hyperref}
\hypersetup{colorlinks=true, linkcolor=blue, citecolor=blue, urlcolor=blue}

\begin{document}

\title{{\bfseries  First-principles study of (Ba,Ca)TiO$_3$ and Ba(Ti,Zr)O$_3$ solid solutions}}

\author{Danila Amoroso}
\affiliation{\footnotesize Physique Th\'eorique des Mat\'eriaux, Q-MAT, CESAM, Universit\'e de Li\`ege (B5), B-4000 Li\`ege, Belgium}
\affiliation{\footnotesize ICMCB, CNRS, Universit\'e de Bordeaux, UPR 9048, F-33600 Pessac, France}
\author{Andr\'es Cano}
\affiliation{\footnotesize ICMCB, CNRS, Universit\'e de Bordeaux, UPR 9048, F-33600 Pessac, France}
\author{Philippe Ghosez}
\affiliation{\footnotesize Physique Th\'eorique des Mat\'eriaux, Q-MAT, CESAM, Universit\'e de Li\`ege (B5), B-4000 Li\`ege, Belgium}

\begin{abstract}

(Ba,Ca)TiO$_3$ and Ba(Ti,Zr)O$_3$ solid solutions are the building blocks of lead-free piezoelectric materials that attract a renewed interest.
We investigate the properties of these systems by means of first-principles calculations, with a focus on the lattice dynamics and the competition between different ferroelectric phases. We first analyze the four parent compounds in order to compare their properties and their different tendency towards ferroelectricity. 
The core of our study is systematic characterization of the binary systems (Ba,Ca)TiO$_3$ and Ba(Ti,Zr)O$_3$ within both the Virtual Crystal Approximation and direct supercell calculations. 
In the case of Ca doping, we find a gradual transformation from $A$-site to $B$-site ferroelectricity due to steric effects that largely determines the behavior of the system. In the case of Zr doping, in contrast, the behavior is eventually dominated by cooperative Zr-Ti motions and the local electrostatics.  
In addition, our comparative study reveals that the specific microscopic physics of these solid sets severe limits to the applicability of the Virtual Crystal Approximation for these systems.
\end{abstract}

\maketitle

\section{Introduction}

The potential interest of BaTiO$_3$-based solid solutions for piezoelectric applications has been known since the sixties,\cite{first_BTZ} although actual piezoelectric devices mainly rely on lead zirconate titanate. However, 
the interest in alternative lead-free solid solutions has been renewed since the 2003 european directive on the restriction of the use of certain hazardous substances in electrical and electronic 
equipment.\cite{eu_law} 
Among all the possible combinations, 
the partial homovalent substitution of Ba by Ca and Ti by Zr in BaTiO$_3$ 
(Ba$_{1-x}$Ca$_{x}$Ti$_{1-y}$Zr$_{y}$ hereafter BCTZ) is of special interest.
In their seminal paper, Liu and Ren\cite{liu_ren} reported a high piezoelectric coefficient of $620~pC/N$ for the ceramic system 
$(1-x)$Ba(Ti$_{0.80}$Zr$_{0.20}$)O$_3$-$x$(Ba$_{0.70}$Zr$_{0.30}$). This has been ascribed to the presence of an special point in the phase diagram in which the tetragonal and rhombohedral ferroelectric phases meet the paraelectric cubic one. Subsequently, 
Keeble \emph{et al.}\cite{maglione2} revisited this phase diagram 
and observed an intermediate ferroelectric orthorhombic phase that also meets the three previous ones. 

As in the case of Pb-based system, the achievement of high piezoelectric response in BCTZ is believed to be linked to the existence of a so-called phase convergence region, 
in which the lack of an energy barrier between different ferroelectric states makes the landscape isotropic and polarization free to rotate\cite{cohen_nature}. 

Despite a great number of experimental studies have been reported,\cite{liu_ren,maglione2,acosta,ranjan} the theoretical investigation of BCTZ compounds remain comparatively very limited and, to the best of our knowledge, no comprhensive study 
based on direct Density Functional Theory (DFT) has been carried out. In order to fill this gap and to clarify the microscopic mechanisms involved 
in the appearence and competition of different polar phases in 
(Ba,Ca)(Ti,Zr)O$_3$ solid solutions, a systematic \emph{ab-initio} investigation is reported here.
Specifically, we report a first-principles study of the structural and dynamical properties of different compositions of 
Ba$_{1-x}$Ca$_{x}$TiO$_3$ and BaTi$_{1-y}$Zr$_{y}$O$_3$ solid solutions and 
related parent compounds. 
We perform calculations using the virtual crystal approximation (VCA) 
and also directly on supercells. The comparison of the results allows us to determine the applicability of VCA for these particular solids solutions. In addition, we also study role of the local atomic arrangements by considering various BaZrO$_3$/$m$BaTiO$_3$ supercells.

The paper is organized as follows. In Sec.~\ref{method} we introduce the technical details of our DFT calculations. Then, 
in Sec.~\ref{parents_compound}, we carefully characterize the origin of the main instabilities via the analysis of the interatomic force constants 
and the energetics of several possible stable and metastable phases in the four parent compounds. 
Next, in Sec.~\ref{bct_solid} and Sec~\ref{btz_solid}, we address the two $A$- and $B$- site substitutions separately in (Ba,Ca)TiO$_3$ and Ba(Ti,Zr)O$_3$ binary-systems, respectively. 
Finally, global discussion and conclusions are reported in Sec.~\ref{discussion} and Sec.~\ref{conclusion}. 

\section{Technical Details }
\label{method}

\subsection{Computational Details}

We use the general framework of the density functional theory (DFT) to compute the structural properties and the electric polarization. 
Dynamical and piezoelectric properties have been calculated using 
Density Functional Perturbation Theory (DFPT)\cite{gonze1}, as implemented in the ABINIT package\cite{abinit}. 
The exchange-correlation potential was evaluated within the generalized gradient approximation (GGA) using the Wu-Cohen (WC) functional\cite{wc-gga} for all simulations.
Optimized Norm-Conserving Pseudopotentials\cite{pseudo} have been employed with the following orbitals considered as the valence states:
$5s$, $5p$ and $6s$ for Ba, $2s$, $2p$ and $4s$ for Ca, $3s$, $3p$, $4s$ and $3d$ for Ti, $4s$, $4p$, $4d$ and $5s$ for Zr and $2s$ and $2p$ for O.
The energy cutoff for the expansion of the electronic wavefunctions has been fixed to 45~Ha. 
Before performing computationally demanding calculations on ordered-alloy supercells by using a standard DFT approach, a first investigation has been addressed
by using the Virtual Crystal Approximation (VCA) within the straightforward combination of the reference atomic pseudopotentials\cite{bellaiche_VCA,phil_VCA1}.
For the optimization of the cubic pervoskite structures, phonons and polarization calculations within VCA, we used a 6x6x6 $k$-points mesh for the Brillouin zone sampling 
for which energy is converged to $0.5~meV$, 
whereas for the optimization of the polar $P4mm$, $R3m$ and $Amm2$ and piezoelectric repsonse calculations different sampling from 8x8x8 to 8x6x6 were used.
For the optimization of the supercells and for the associated phonons calculations we used the 8x8x8 $k$-mesh. For all the tetragonal superlattices we used 8x8x6 $k$-mesh sampling. 
In order to allow comparison between different structures, energy will be reported in $meV$/f.u. (i.e. per 5 atoms).
The $q$-points for the phonon dispersion curves and IFCs of the four pure compounds included $\Gamma$, $X$, $M$, $R$ and the $\Lambda$ point halfway from $\Gamma$ to $R$ of the 
simple cubic Brillouin zone.

\subsection{Structure of the parent compounds}

The optimized the lattice parameters of the cubic structure are reported in Table~\ref{eigen_bulk}. These values are used in Sec. \ref{phonon_bulks_ifc} for the calculation of the phonon dispersion curves and interatomic force constants. The obtained values are in excellent agreement with experimental data for BaTiO$_3$ and BaZrO$_3$ \cite{BTO_cubic1, BZO_cubic1}
(within 1$\%$), whereas for CaTiO$_3$ the understimation is of about 2$\%$ \cite{CTO_cubic1}. The larger error for CaTiO$_3$ can be assigned to the fact that 
the $0~K$ DFT results are compared to measurements at 1720 K.
For CaZrO$_3$ the optimized lattice parameter overstimates the experimental value \cite{CZO_cubic1} of about 2$\%$. 

Atoms in the $ABO_3$ perovskite structure are labeled according to Fig~\ref{atoms_scheme}.

\subsection{Supercell structures}
\label{supercell_description}

\begin{figure}[t]
\centering
\includegraphics[width=5.5cm]{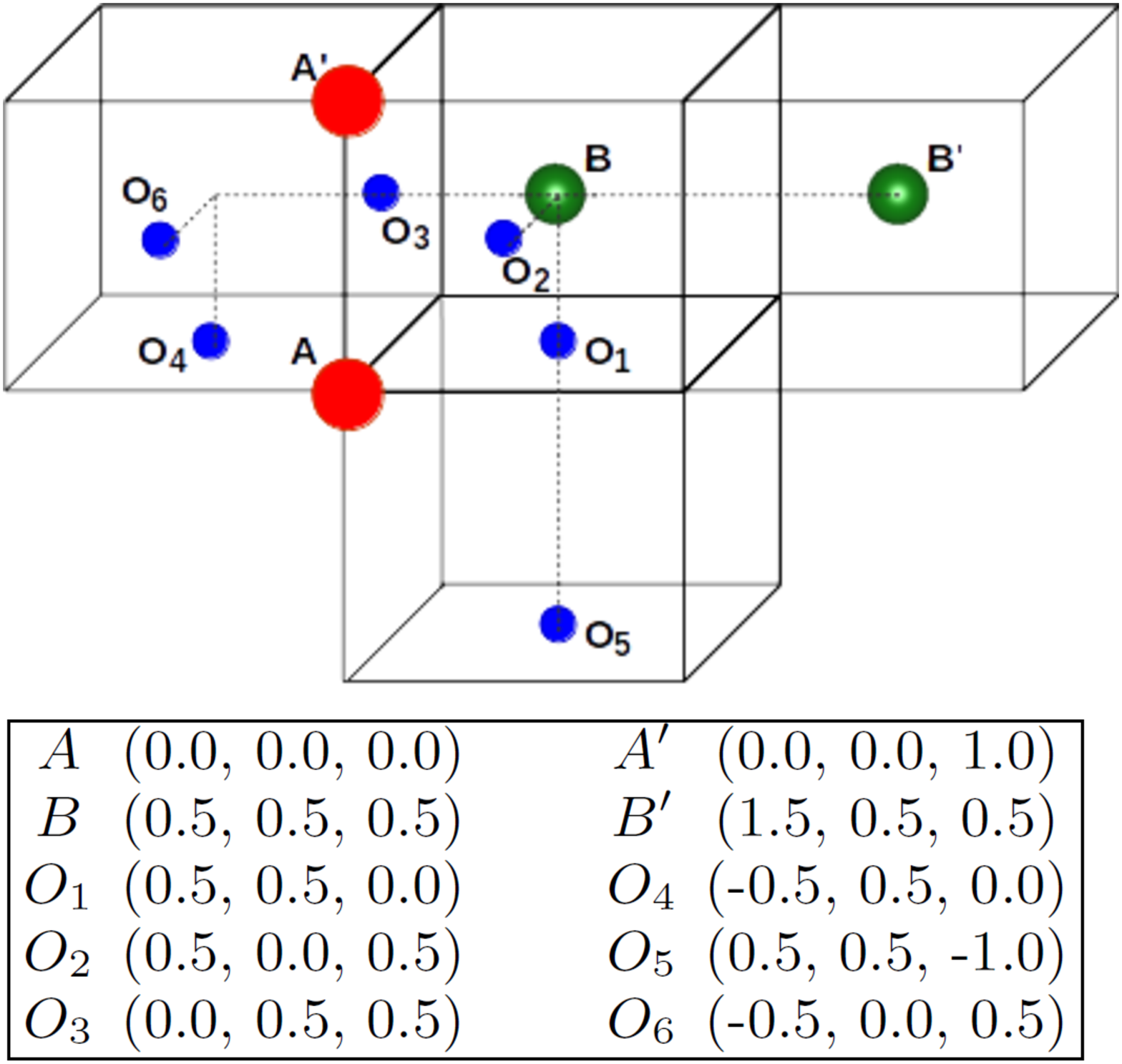}
\caption{\footnotesize Top: schematic 3D-view of atoms. Bottom: positions in reduced coordinates of the atoms in the perovskite structure.}
\label{atoms_scheme}
\vspace{0.4cm}
\includegraphics[width=\columnwidth]{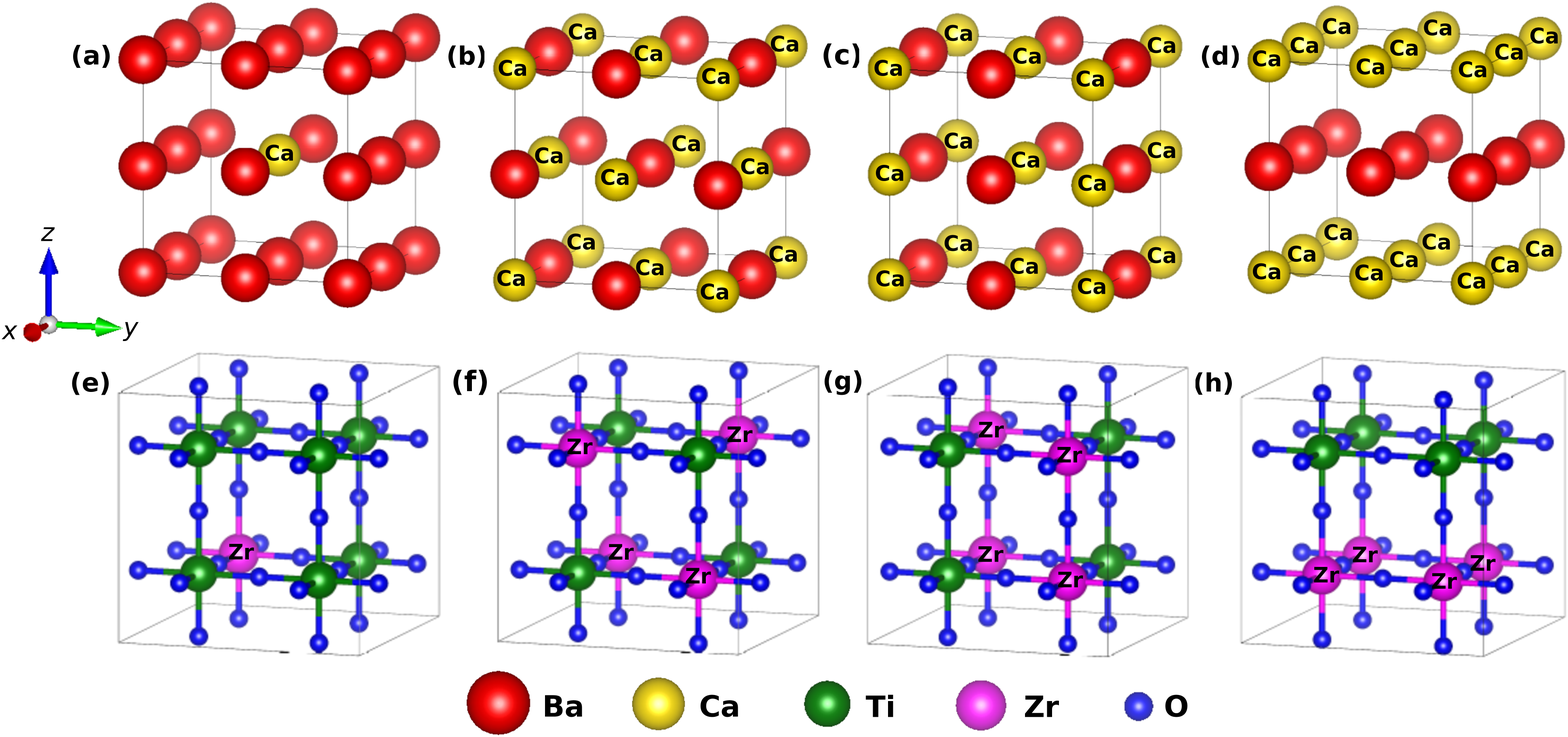}
\caption{\footnotesize Schematic representation of different atomic ordering in 2x2x2 supercells for the two investigated solid solutions.
Top: sublattice of $A$ cations in Ba$_{1-x}$Ca$_{x}$TiO$_3$. Bottom: sublattice of $B$ cations and oxygens in BaTi$_{1-y}$Zr$_{y}$O$_3$. 
Ca and Zr atoms are labelled to help the visualization.
(a,e) $x,y=0.125$. (b,f) rocksalt order, (c,g) columnar order along the $z$ axis and (d,h) layered order perpendicular to $z$ axis for $x,y=0.50$.} 
\label{super_12}
\end{figure}

\begin{table}[b]
\centering
\footnotesize
\begin{tabular}{|c|c|c||c|}
\hline
\multicolumn{2}{|c|}{Structure} & {\bf Ba$_{0.50}$Ca$_{0.50}$TiO$_3$} & {\bf BaTi$_{0.50}$Zr$_{0.50}$O$_3$}\\
\hline
          & columnar & $a=b=5.535$~\AA  &  $a=b=5.752$~\AA  \\
$P4/mmm$ & [001]  & $c=3.899$~\AA    &  $c=4.104$~\AA   \\[4pt]
          & layered & $a=b=3.911$~\AA  &  $a=b=4.098$~\AA  \\
          & [110]  & $c=7.803$~\AA    &  $c=8.125$~\AA   \\
\hline
\multicolumn{2}{|c|}{$Fm\bar3m$} & $a=b=c=6.129~$~\AA & $a=b=c=8.150$~\AA\\
\hline
\end{tabular}
\caption{\footnotesize Optimized lattice parameters (in \AA) for Ba$_{0.50}$Ca$_{0.50}$TiO$_3$ and BaTi$_{0.50}$Zr$_{0.50}$O$_3$ solid solutions for
the tetragonal $P4/mmm$ and the cubic $Fm\bar3m$ structures.
For the tetragonal $P4/mmm$ we report values related to the reduced 10-atoms cell and relations with respect to the 2x2x2 supercell are:
(chains) a=b=a$_{(2x2x2)}/\sqrt2$ and c=c$_{(2x2x2)}/2$; (planes) a=b=a$_{(2x2x2)}/2$ and $c$=$c_{(2x2x2)}$. Structural transformation have been done via
the FINDSYM tool\cite{isotropy}. }
\label{lattice_parameter_50}
\end{table}

We have considered different supercells describing the Ca and Zr doping of BaTiO$_3$ and performed DFPT calculations on the corresponding relaxed high-symmetry
paraelectric structures.

For Ba$_{7/8}$Ca$_{1/8}$TiO$_3$ and BaTi$_{7/8}$Zr$_{1/8}$O$_3$, we have used the smallest reference structure which is the
cubic $Pm\bar{3}m$ ($O^{h}_{1}$) with 40 atoms. The GGA-WC optimized lattice constants are $7.917~$\AA~ and $8.003~$\AA~ respectively (Fig.~\ref{super_12}).

For Ba$_{0.50}$Ca$_{0.50}$TiO$_3$ and BaTi$_{0.50}$Zr$_{0.50}$O$_3$
three distinct geometric arrangements
of the 40-atoms cell were considered:
columns of same cations along the [001] direction, layers of same cations parallel to the [110] planes and rocksalt configuration, Fig.~\ref{super_12}.
The paraelectric reference of the two first ``ordered" structures are characterized by a tetragonal symmetry within the $P4/mmm$ ($D^{4h}_{1}$) space-group and they can be reduced by symmetry to
10-atoms cell, whereas the rocksalt configuraton has the cubic symmetry within the $Fm\bar{3}m$ ($O^{h}_{5}$) space-group.
Structural relaxation have been done on the 2x2x2 supercells (40 atoms) for each geometrical ordering, whereas DFPT calculations 
and structural relaxation of lower symmetry structures 
have been performed on the reduced 10-atoms cell for the $P4/mmm$ structures and on the 40-atoms cell for the $Fm\bar{3}m$ one.
The optimized lattice parameters of the high-symmetry references are reported in Table~\ref{lattice_parameter_50}. 
Figures have been produced by using the VESTA package\cite{vesta}. 

\section{Parent compounds}
\label{parents_compound}

We start our study by considering individually the BaTiO$_3$, CaTiO$_3$, CaZrO$_3$ and BaZrO$_3$ parent compounds. 
The dynamics and the energetics of these systems is fundamental to understand the key features of corresponding solid solutions.

BaTiO$_3$ is one of the most studied perovskite, both from the theoretical \cite{devonshire, cochran, BTO_eff, phil2} and 
experimental \cite{BTO_ortho, BTO_rhombo, BTO_kwei} points of view. This perovskite is characterized by a tolerance factor $t$ greater than 1, $t\simeq 1.06$\cite{tolerance_factor1}, 
that 
allows to predict a polar distorted ground-state \cite{tolerance_factor1}. In fact, while stable at high temperature in the centrosymmetric cubic ($Pm\bar3m$) phase, it 
undergoes \cite{BTO_kwei} ferroelectric structural phase transitions to a tetragonal ($P4mm$) structure at $\simeq 393~K$, to a orthorhombic ($Amm2$) phase at $\simeq 278~K$ 
and to a rhombohedral ($R3m$) ground-state at $183~K$. 

CaTiO$_3$, on the contrary, has a tolerance factor $t$ smaller than 1, $t\simeq 0.97$\cite{tolerance_factor1}. Accordingly, this compound displays a non-polar orthorhombic ($Pnma$) ground-state.
 It  exhibits at least two observed phase transitions at high temperatures: from a cubic ($Pm\bar3m$) to a tetragonal phase ($I4/mcm$) at $\simeq 1634~K$ and from a 
tetragonal ($I4/mcm$) to the orthorhombic ground-state ($Pnma$) 
at $\sim 1486$ K \cite{CTO_cubic1, CTO_kennedy, CTO_tem}.

CaZrO$_3$, that has $t\simeq 0.91$, was observed only in two different structures: 
the high temperature cubic ($Pm\bar3m$) form and its orthorhombic ($Pnma$) ground-state with the transition temperature at $\simeq 2170~K$.\cite{CZO_stoch, CZO_hou, CZO_du}

BaZrO$_3$, with 
$t\simeq 1.00$\cite{tolerance_factor1}, is 
experimentally not known to undergo any structural phase transition 
and remains in the paraelectric cubic phase down to $2$ K. \cite{BZO_cubic1}

In Section \ref{phonon_bulks_ifc}, we provide a comparison of the dynamical properties (phonon dispersion curves and interatomic force constants) 
of the parent compounds in their cubic phase. 
Then, in Section \ref{bulk_energy}, we compare the energetics of various metastable phases of lower symmetry arising from the condensation
of individual and combined distortions related to the unstable phonon modes identified in Sec.~\ref{phonon_bulks_ifc}.

\subsection{Phonon dispersion curves and Interatomic Force Constants}
\label{phonon_bulks_ifc}

\begin{table}[t]
\centering
\footnotesize
\begin{tabular}{|c|c|c|c|c|}
\hline
 &  {\bf BaTiO$_3$} & {\bf CaTiO$_3$}  & {\bf BaZrO$_3^{**}$} & {\bf CaZrO$_3$} \\
\hline
\multicolumn{5}{|c|}{Lattice parameter $a_{cell}$ (\AA)}  \\
\hline
Present  &  3.975    & 3.840     &  4.184     & 4.099 \\
Exp.     &  4.003\cite{BTO_cubic1} & 3.897\cite{CTO_cubic1} & 4.191\cite{BZO_cubic1} & 4.020\cite{CZO_cubic1} \\
\hline
Atom  & \multicolumn{4}{c|}{Born effective charge $Z^{*}$}  \\
\hline
$A^{(+2)}$             &  2.751 & 2.575 & 2.732    & 2.623  \\
$B^{(+4)}$             & 7.289 & 7.188 & 6.099    & 5.903  \\
O$^{(-2)}_{\parallel}$ & -5.756 & -5.730  & -4.808 & -4.862  \\
O$^{(-2)}_{\perp}$     & -2.142 & -2.017 & -2.012  & -1.832  \\
\hline
\hline
\multicolumn{5}{|c|}{Phonon eigendisplacements} \\
\hline
\multicolumn{5}{|c|}{$F_{1u}(TO_1)$} \\
\hline
$\omega$ (cm$^{-1}$) & 183.45i  & 136.21i  & 96.39  & 179.91i   \\
$\bar{Z}^{*}$ & 9.113 & 6.453 & 3.954 & 4.455 \\
\hline
$A$         & +0.0012 & +0.0950 & +0.0561 & +0.1120 \\
$B$         & +0.0978 & +0.0298 & -0.0337 & +0.0002 \\
O$_1$     & -0.1480 & -0.0763 & -0.0687 & -0.0390 \\
O$_{2/3}$ & -0.0774 & -0.1254 & -0.1103 & -0.1215 \\
\hline
\multicolumn{5}{|c|}{$F_{1u}(TO_2)$} \\
\hline
$\omega$ (cm$^{-1}$) & 176.91  & 181.36  & 193.26  & 202.42    \\
$\bar{Z}^{*}$ & 1.937 & 5.344 & 5.784 & 3.268 \\
\hline
$A$         & +0.0547  & -0.0922  & -0.0228 & -0.0823 \\
$B$         & -0.0800 & +0.1116 & +0.0781  & +0.0698 \\
O$_1$     & -0.0715   & -0.0189  & -0.0353 & -0.0039 \\
O$_{2/3}$ & -0.0793   & -0.0421  & -0.1068 & -0.0940 \\
\hline
\multicolumn{5}{|c|}{$F_{1u}(TO_3)$} \\
\hline
$\omega$ (cm$^{-1}$) & 468.91  & 589.62 & 503.07  & 610.29    \\
$\bar{Z}^{*}$ & 1.281 & 4.269 & 3.777 & 4.601 \\
\hline
$A$         & -0.0012 & +0.0083 & +0.0024 & -0.0065 \\
$B$         & +0.0253  & +0.0135  & +0.0140 & -0.0225  \\
O$_1$     & +0.1767    & -0.2213  & -0.2315 & +0.2353 \\
O$_{2/3}$ & -0.1212     & +0.0801 & +0.0653 & -0.0454 \\
\hline
\end{tabular}
\begin{tabular}{l}
\footnotesize $^{**}$ BaZrO$_3$ has no instabilities at $\Gamma$, whereas the $TO_1$ mode\\
\footnotesize ~~~ is unstable for BaTiO$_3$, CaTiO$_3$ and CaZrO$_3$. \\
\end{tabular}
\caption{\footnotesize Relaxed lattice parameter $a_{cell}$ (in \AA)~at which DFPT calculations have been performed in the cubic perovskite structure for the four $ABO_3$ compounds.
Experimental values are also reported for comparison.
Born effective charges, normalized phonon eigendisplacements (in $a. u.$) and mode effective charges for the $F_{1u}(TO)$ modes at $\Gamma$ from DFPT calculations
on the optimized cubic phase. The latter quantity is defined as in Ref.\cite{gonze1}, $\bar{Z}^{*}_{TO}=\left|\frac{\sum_{k,\beta}Z^{*}_{k,\alpha\beta}\eta^{TO}_{k,\beta}}{\sqrt{<\eta^{TO}|\eta^{TO}>}}\right|$}
\label{eigen_bulk}
\end{table}

\begin{figure*}[t]
\centering
\includegraphics[width=17cm]{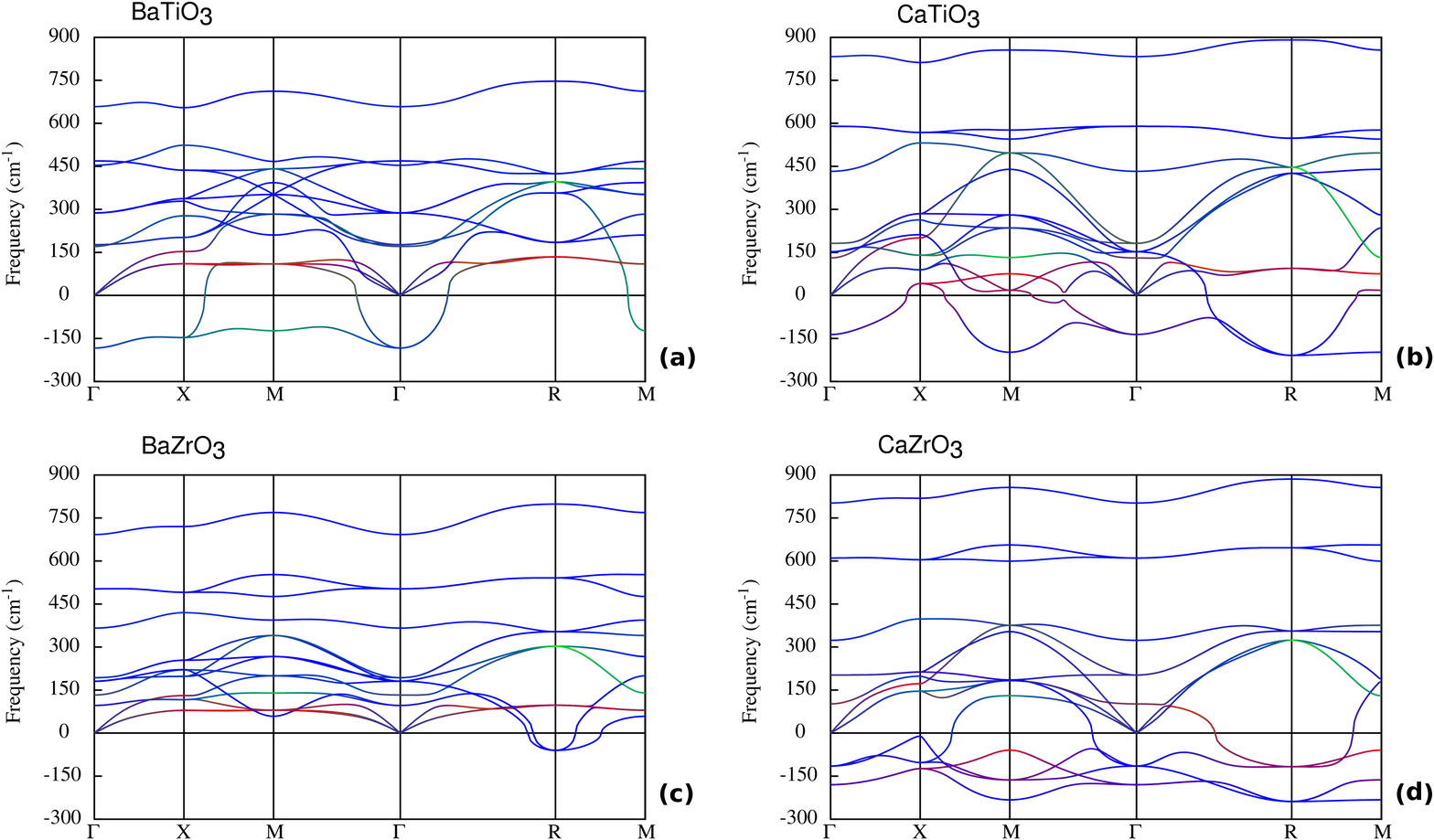}
\caption{\footnotesize Calculated phonon dispersion curves of BaTiO$_3$ (a), CaTiO$_3$ (b), BaZrO$_3$ (c) and CaZrO$_3$ (d) along different high-symmetry lines in 
the simple cubic Brillouin zone. Negative values of frequencies refer to imaginary frequencies ($\omega$ are in $cm^{-1}$). The different colors indicate the different atomic 
contributions in the corresponding eigenvectors as in Ref.\cite{phil1} (red for the $A$ atom, green for $B$ atom and blue for $O$ atoms).}
\label{bulk_phon}
\end{figure*}

The identification of imaginary phonon frequencies and the corresponding displacement patterns allow us to pinpoint the main instabilities behind the structural phase transitions in our systems.
Thus, we considered the cubic reference structure and we computed the phonon dispersion curves along selected high-symmetry lines in the simple Brillouin zone. 
The results are shown in Fig.~\ref{bulk_phon}. 
The presence of imaginary phonon frequencies (shown as negative values in Fig.~\ref{bulk_phon}) reveals the structural instabilities of the cubic phase.
The nature of the corresponding transition is usually determined by the character of the main unstable modes and the related eigendisplacement vectors indicate the atomic displacements
that spontaneously appear to reach the most stable configuration.

We complement this analysis with the calculation of the interatomic force constanst in real space. These constans are defined as 
$C_{k\alpha,k'\beta}(l,l')=\partial^{2}E_{tot}/\partial \tau^{l}_{k\alpha}\partial \tau^{l'}_{k'\beta}$, where $E_\text{tot}$ is the total energy of a period crystal 
and $\tau_{k\alpha}^l$ is the displacement of the atom $k$ in the cell $l$ along direction $\alpha$ from its equilibrium position.
For a pair of distinct atoms, the IFC can be interpreted as minus the harmonic spring constant between them,
so that negative values correspond to stable interactions. For the ``on-site'' forces, on the contrary, positive values correspond to stable interactions.
For a more detailed description of the physical meaning see Refs.~\cite{phil1,gonze1,gonze2}. 
In our analysis, we further separate the contribution of the dipole-dipole interaction (DD) from that of
the short-range forces (SR) in order to identify the key mechanisms that lead the system to display or not specific phonon instabilities.

In the following we address a systematic description of the phonon dispersion curves reproduced in Fig.~\ref{bulk_phon}. 

\subsubsection{BaTiO$_3$}
\label{BTO_phonon}

Ferroelectricity in BaTiO$_3$ is known to be related to the Ti off-centering driven by an unstable polar mode at $\Gamma$ \cite{phil2, phil1}. 
This ferroelectric (FE) instability is such that it expand over the entire $\Gamma-X-M$ planes of the simple cubic Brillouin zone, as can be seen in Fig.~\ref{bulk_phon}(a). 
While each atom is at a position stable against individual displacements (Table~\ref{sfc_bulk}),
the origin of this distortion has to be primarily ascribed to the destabilizing Ti-O interaction, reflected in the positive value of the interatomic force constant along the 
bond's direction (Table~\ref{ifc_bulk_local}). An additional contribution comes from the strong interaction of subsequent Ti-atoms along the $B$-$B'$ chain direction compared to the small 
value in the transverse component. 
These anisotropic couplings give rise to dispersion of the unstable mode almost flat in the $\Gamma-X-M$ plane and highly dispersive along $\Gamma-R$. This
reflects the chain-like nature of the instability in real space. The polar distortion requires cooperative atomic displacement along Ti-O chains\cite{phil2}.
Furthermore, the negligible contribution of Ba-atom to the ferroelectric distortion, has to be ascribed to its sizable on-site force constant and very weak Ba-O$_1$ coupling.

In Table~\ref{eigen_bulk}, we also compare the eigendiplacements associated to the unstable $F_{1u}(TO_1)$ mode to those of the 
stable $F_{1u}(TO_2)$ and $F_{1u}(TO_3)$ ones and we report the related mode effective charges, as described in Ref.\cite{gonze1}. 
The mode effective charge is giant for the unstable mode in comparison to the others\cite{polarity, phil_polarity}.
 In fact, the very anomalous Born effective charges on Ti and O$_{\parallel}$ 
combined through the specific pattern of eigendiplacements associated to the $TO_1$ mode in order to produce a large spontaneous polarization, while for the $TO_2$ and $TO_3$ modes, 
the motions are coupled so that Ti and O generate polarizations that partly compensate.

\subsubsection{CaTiO$_3$} 
\label{CTO_phonon}

CaTiO$_3$ displays two main non-polar instabilities at the $R$ and $M$ points of the cubic Brillouin zone
related to antiferrodistortive (AFD) motions ($\omega \simeq 209i~cm^{-1}$ and $\omega \simeq 198i~cm^{-1}$, respectively). 
These correspond to cooperative rotations of oxygen octahedron, around the $B$-atoms, with consecutive octahedron along the rotation axis being in-phase at $M$ 
($a^0a^0c^+$ in Glazer's notation \cite{glazer1}) and anti-phase at $R$ ($a^0a^0c^-$ ).
As such, the $M$-instability appearing in the phonon spectrum is a continuation of the instability at $R$, 
while in BaTiO$_3$ it is a continuation of the polar instability at $\Gamma$. A detailed description of such AFD-instabilities is reported in Ref.\cite{safari}.

In addition, there is an unstable mode at the $\Gamma$-point that is also polar. 
This FE instability is now restricted to a region around the $\Gamma$-point highlighting a larger and more isotropic ferroelectric correlation volume.
This mode enables the condensation of a polar distortion. 
However, the character of the corresponding ferroelectricity is rather different 
compared to that in BaTiO$_3$ where it is dominated by the Ti displacements ($B$-site ferroelectricity). 
In fact, in CaTiO$_3$, this instability turns out to be dominated by the Ca displacements ($A$-type ferroelectricity) as can be seen from the eigendisplacements displayed in  
Table~\ref{eigen_bulk} and the red color of the corresponding phonon in Fig.~\ref{bulk_phon}(b).
Despite this important involvement of the $A$-cation in the polar distortion, 
its Born effective charge is not strongly anomalous. 
This has to be ascribed to fact that the polar distortion of Ca is driven by a steric effect. 
In fact, we already mentioned that CaTiO$_3$ has a tolerance factor less than one. Therefore, the small size of the Ca ionic radius allows its distortion in the cubic perovskite structure.
Consequently, the involvement of the oxygens in the distortion is inverted with respect to the BaTiO$_3$ case, as 
the O$_{2/3}$ lying on the plane perpendicular to the direction of the distortion are now more involved than the apical oxygen (see Table~\ref{eigen_bulk}).

These results can be rationalized by looking at the effect of the substitution of Ba by Ca on the on-site and interatomic force constants.
In fact, the ``on-site" force constant of Ca as well as the $A$-$A'$ interaction are 
significantly smaller than the corresponding ones in BaTiO$_3$, while the ``on-site" force constant of Ti is increased.
Additionally, the destabilizing $A$-O$_1$ interaction
 ($xx'$, Table~\ref{ifc_bulk_local}) become significantly positive in agreement with the opposite direction
of the respective atomic eigendisplacements (Table~\ref{eigen_bulk}), whereas along the Ti-O$_1$ chain a stronger repulsive interaction prevents an important participation of
the $B$-cation to the polar distortion. 
 
These previous observations can be also related to the concurrent appearance of AFD distortions in the system.
Although the IFCs between the reported pairs of oxygen are remarkably similar in BaTiO$_3$ and CaTiO$_3$ (and also in BaZrO$_3$ and CaZrO$_3$ as we will see below), 
the oxygens tilting is favored by the fast decrease of the oxygen ``on-site'' forces (linked to the destabilizing $A$-O$_1$ interaction) 
in the directions perpendicular to the $B$-O chains and to the increasing stiffness
in the parallel directioni (Table~\ref{sfc_bulk}).
Moreover, the phonon dispersion curves appear
substantially flat along $R-M$ suggesting the absence of coupling between the oxygens in different planes, but, as can be seen in Tables~\ref{ifc_bulk_local}
and~\ref{ifc_bulk_cart}, the transverse interactions are far to be negligible. As proposed in Ref.\cite{phil1}, this could be due to the joint action
of the $A$-O coupling with a compensation between different interplane interaction of the oxygens (see also Ref.\cite{bellaiche_AO}). 

\subsubsection{CaZrO$_3$} 
\label{CZO_phonon}

CaZrO$_3$ exhibits much more intricate phonon branches than CaTiO$_3$.
However, the dynamical properties of both compounds show some similarities. 
As in CaTiO$_3$, the strongest instabilities of CaZrO$_3$ are at the $R$ and $M$ points of the Brillouin zone 
and associated to AFD oxygens rotations ($\omega \simeq 238i~cm^{-1}$ and $\omega \simeq 233i~cm^{-1}$, respectively). Also, the dispersion curve along 
 $R-M$ appears completely flat. 

Nevertheless, the lowest polar instability is no more confined at $\Gamma$, but the polar-antipolar instability extends all over the cubic Brillouin zone. 
This reflects a FE instability very localized in real space.

In CaZrO$_3$, the distortion is fully driven by the $A$-cation (Ca) and the O$_{2/3}$ anions. 
In fact, Zr on the $B$-site does not participate to the polar distortion with also subsequent reduced participation of the apical O$_1$ (see Table~\ref{eigen_bulk}-($TO_1$ mode)). 
Finally, the dynamical effective charges of both Zr and the related O appear less anomalous than in the titanates resulting in 
a lower value of the mode effective charge. 

As before, these dynamical properties can be understood in terms of IFCs. In fact, the ``on-site'' force associated to Ca and the in-plane ``on-site'' force of O are drastically reduced 
while for Zr it is strengthened by making the $A$-cation and the O$_{2/3}$ anions highly unstable and the $B$-cation almost fixed with respect to their high-symmetry positions. 
Accordingly, the Ca-O$_1$ interatomic force constants are largely dominated by the dipole-dipole interaction, 
while the Zr-O$_1$ interaction is strongly repulsive. 
The interplay between all these features allows the understanding of the strong $A$-site driven character of the instabilities.
In fact individual Ca displacements are nearly unstable (Table~\ref{sfc_bulk}).

A proper comparison of CaZrO$_3$ an CaTiO$_3$ with respect to PbZrO$_3$ and PbTiO$_3$ respectively, can be done via the lattice dynamics analysis presented in Ref.\cite{phil1}.

\subsubsection{BaZrO$_3$}
\label{BZO_phonon}

Starting from the analyzed dynamics in BaTiO$_3$ and CaZrO$_3$, the dynamical properties of BaZrO$_3$ are fairly predictable. 
The substitution of the Ti atoms with Zr on the $B$-site joined to
the presence of Ba on the $A$-site leads to the stability of the cubic phase, also confirmed by steric arguments, since the tolerance factor is close to 1.
In fact, unlike the previous perovskite systems, the phonon spectrum of BaZrO$_3$ shows only a very weak instability at the $R$-point 
with associated phonon frequency $\omega \simeq 60.2i~cm^{-1}$ and no unstable polar modes. 

By looking at the specific quantities reported in Table~\ref{eigen_bulk}, 
the softest polar mode ($TO_1$) displays the smallest polarity ($\bar{Z}^{*}$) with respect to the corresponding modes in the other compounds. 
Moreover, unlike Ti in BaTiO$_3$, a tendency of Zr to decrease the spontaneous polarization results from the specific combination of the associated pattern of distortion with the dynamical charges. 
Conversely, the second stable polar mode ($TO_2$), even if stiffer, displays bigger polarity because of the additive contribution of Zr and O. 

This behaviour can be justified by means of the force constants. In fact, the $B$-cation experiences an increased ``on-site'' term as well as O in its transversal direction
with respect to the case of BaTiO$_3$.
Accordingly, the Zr-O$_1$ is strongly dominated by the short-range forces.
On the other hand, the $A$-cation shows an ``on-site'' force constant in between BaTiO$_3$ and CaTiO$_3$, 
but the $A$-O interaction is still too weak to destabilize Ba (see Tables~\ref{sfc_bulk} and \ref{ifc_bulk_local}). \\

For all the four compounds, the interatomic force constants associated to the $A-B$ cations and to the O-O interactions have the same sign and
 order of magnitude. This trend points out the key role played by the on-site force constants in generating different kinds of instabilities according
to the environment that each atom experiences in the different perovskites. In fact, changes in the
$A$-O and $B$-O interactions (Tables~\ref{sfc_bulk} and~\ref{ifc_bulk_local}) are primarily associated to changes of that values.

\begin{table}[b]
\centering
\begin{tabular}{|c|c|c||c||c||c|}
\hline
Atom  & Direction    &  {\bf BaTiO$_3$}   & {\bf CaTiO$_3$}  & {\bf BaZrO$_3$}  &  {\bf CaZrO$_3$} \\
\hline
$A$   &  $\small{xx=yy=zz}$  &  +0.0893  &  +0.0269  &  +0.0550  &  +0.0099  \\
\hline
$B$    & $\small{xx=yy=zz}$  &  +0.1635  &  +0.2236  &  +0.2198  &  +0.2510  \\
\hline
$O$   &  $\small{xx=yy}$     &  +0.0711  &  +0.0432  &  +0.0396  &  +0.0171  \\
      &    $\small{zz}$      &  +0.1454  &  +0.2624  &  +0.2135  &  +0.2909  \\
\hline
\end{tabular}
\caption{\small ``On-site'' force constants (in Ha/Bohr$^2$) related to different atoms in the cubic phase of the four bulk compounds. 
A positive value means that the atomic position is stable against isolated displacement of the considered atom. }
\label{sfc_bulk}
\end{table}

\begin{table*}[t]
\footnotesize
\centering
\begin{tabular}{|cc|c|c|c||c|c|c||c|c|c||c|c|c|}
\hline
     &  & \multicolumn{3}{c||} {\bf BaTiO$_3$} & \multicolumn{3}{c||}{\bf CaTiO$_3$} &  \multicolumn{3}{c||}{\bf BaZrO$_3$} & \multicolumn{3}{c|}{\bf CaZrO$_3$} \\
Atoms &             &   Total &   DD    &   SR  &   Total &   DD    &   SR  &   Total &   DD    &   SR  &   Total &   DD    &   SR  \\
\hline
$A-A'$ & ($\parallel$) & -0.0115 & -0.0054 & -0.0061 & -0.0085 & -0.0056 & -0.0029  & -0.0094  & -0.0062 & -0.0032 & -0.0071 & -0.0064 & -0.0007  \\ 
       & ($\perp$)     & +0.0039 & +0.0027 & +0.0012 & +0.0040 & +0.00281 & +0.0012 & +0.0130 & +0.0106  & +0.0024 & +0.0042  & +0.0032 & +0.0010   \\
\hline
$B-B'$ & ($\parallel$) & -0.0693 & -0.0379 & -0.0314 & -0.0788 & -0.0438 & -0.0350 & -0.0564 & -0.0308 & -0.0256 & -0.0593 & -0.0323 & -0.0270 \\
       & ($\perp$)     & +0.0078 & +0.090 & -0.0111 & +0.0084 & +0.0219 & -0.0135 & +0.0071 & +0.0154 & -0.0083 & +0.0065 & +0.0162 & -0.0097 \\
\hline
$B-O_1$ & ($\parallel$) & +0.0037 & +0.2394 & -0.2357 & -0.0382 & +0.2794 & -0.3176 & -0.0409 & +0.1940 & -0.2349 & -0.0719 & +0.2129 & -0.2847 \\
        & ($\perp$)     & -0.0203 & -0.0445 & +0.0243 & -0.0184 & -0.0492 & +0.0308 & -0.0166 & -0.0406 & +0.0240 & -0.0148 & -0.0401 & +0.0253 \\
\hline
$A-B$   & ($\parallel$) & -0.0298 & -0.0220 & -0.0078 & -0.0266 & -0.0242 & -0.0025 & -0.0281 & -0.0212 & -0.0069 & -0.0244 & -0.0221 & -0.0023 \\
        & ($\perp$)     & +0.0139 & +0.0110 & +0.0029 & +0.0150 & +0.0121 & +0.0029 & +0.0130 & +0.0106 & +0.0024 & +0.0133 & +0.0111 & +0.0023 \\
\hline
$A-O_1$ & ($xx'$) & -0.0022 & +0.0119 & -0.0140 & +0.0108 & +0.0125 & -0.0017 & +0.0058 & +0.0129 & -0.0070 & +0.0141 & +0.0126 & +0.0015 \\
        & ($yy'$) & -0.0042 & -0.0059 & +0.0017 & -0.0055 & -0.0062 & +0.0007 & -0.0051 & -0.0064 & +0.0013 & -0.0056 & -0.0063 & +0.0007 \\
        & ($zz'$) & -0.0111 & -0.0160 & +0.0049 & -0.0116 & -0.0177 & +0.0061 & -0.0105 & -0.0154 & +0.0049 & -0.0107 & -0.0167 & +0.0060 \\
\hline
$O_1-O_4$ & ($xx'$) & -0.0020 & -0.0033 & +0.0013 & -0.0022 & -0.0035 & +0.0012 & -0.0019 & -0.0034 & +0.0014 & -0.0020 & +0.0012 & +0.0078 \\
          & ($yy'$) & +0.0018 & +0.0016 & +0.0002 & +0.0014 & +0.0017 & -0.0003 & +0.0018 & +0.0017 & +0.0001 & +0.0012 & +0.0016 & -0.0003 \\
          & ($zz'$) & +0.0093 & +0.0118 & -0.0025 & +0.0107 & +0.0139 & -0.0032 & +0.0070 & +0.0096 & -0.0026 & +0.0078 & +0.0110 & -0.0032 \\
\hline
$O_1-O_5$ & ($xx'$) & -0.0004 & +0.0016 & -0.0020 & -0.0009 & +0.0017 & -0.0026 & -0.0009 & +0.0017 & -0.0026 & -0.0013 & +0.0016 & -0.0028 \\
          & ($yy'$) & -0.0004 & +0.0016 & -0.0020 & -0.0009 & +0.0017 & -0.0026 & -0.0009 & +0.0017 & -0.0026 & -0.0013 & +0.0016 & -0.0028 \\
          & ($zz'$) & -0.0339 & -0.0236 & -0.0102 & -0.0379 & -0.0278 & -0.0101 & -0.0345 & -0.0191 & -0.0154 & -0.0372 & -0.0219 & -0.0153 \\
\hline
\end{tabular}
\caption{\small Interatomic force constants in (Ha/Bohr$^2$) between different pairs of atoms in their local coordinates system, $xx'~(\parallel)$, $yy'~(\perp)$ and $zz'~(\perp)$ 
for $ABO_3$ bulk compounds.
Transverse ($\perp$) directions for some atomic pairs are degenerate. The two different dipole-dipole (DD) and short-range (SR) contributions also reported.}
\label{ifc_bulk_local}
\centering
\begin{tabular}{|c|cc|}
\hline
Atoms & & \\
\hline
& \bf {BaTiO$_3$}  & \bf {CaTiO$_3$} \\
 & $\begin{pmatrix}
+0.0038 & 0.0000 & 0.0000 \\ 
0.0000 & -0.0091  & +0.0284 \\
0.0000 & +0.0126  & -0.0091
\end{pmatrix} $               &      $\begin{pmatrix}  
+0.0057 & 0.0000 & 0.0000 \\
0.0000 & -0.0120 & +0.0322 \\
0.0000 & +0.0177 & -0.0120
\end{pmatrix} $          \\
$O_1-O_2$ & & \\
& \bf {BaZrO$_3$} & \bf {CaZrO$_3$} \\
 &      $\begin{pmatrix} 
+0.0039 & 0.0000  & 0.0000   \\
0.0000 & -0.0060 & +0.0234 \\
0.0000 & +0.0110 & -0.0060  
\end{pmatrix} $           &      $\begin{pmatrix}
+0.0051 & 0.0000 & 0.0000 \\
0.0000 & -0.0071 & +0.0249 \\
0.0000 & +0.0137 & -0.0071 
\end{pmatrix} $ \\
\hline
& \bf {BaTiO$_3$} & {\bf CaTiO$_3$} \\
 & $\begin{pmatrix}
-0.0007 & -0.0007 & +0.0013 \\ 
-0.0013 & +0.0014 & +0.0025  \\
+0.0007 & +0.0007  & +0.0014
\end{pmatrix} $               &      $\begin{pmatrix}  
-0.0007 & -0.0006 & +0.0017 \\
-0.0017 & +0.0015 & +0.0029 \\
+0.0006 & +0.0005 & +0.0015
\end{pmatrix}$          \\
$O_1-O_6$ & & \\
& {\bf BaZrO$_3$} & {\bf CaZrO$_3$} \\
 &      $\begin{pmatrix} 
-0.0077 & -0.0007 & +0.0013  \\
-0.0013 & +0.0012 & +0.0021 \\
+0.0007 & +0.0006 & +0.0012 
\end{pmatrix}$           &
$\begin{pmatrix}
-0.0007 & -0.0005 & +0.0016 \\
-0.0016 & +0.0013 & +0.0023 \\
+0.0005 & +0.0004 & +0.0013 
\end{pmatrix}$ \\
\hline
\end{tabular}
\caption{\small Interatomic force constant matrix in (Ha/Bohr$^2$) between other pairs of Oxygen in the Cartesian coordinates system. Rows and columns of the matrices correspond, respectively,
 to $x, y$ and $z$ directions.}
\label{ifc_bulk_cart}
\end{table*}

\subsection{Energetics of metastable phases}
\label{bulk_energy}

\begin{figure}[h]
\centering
\includegraphics[width=\columnwidth]{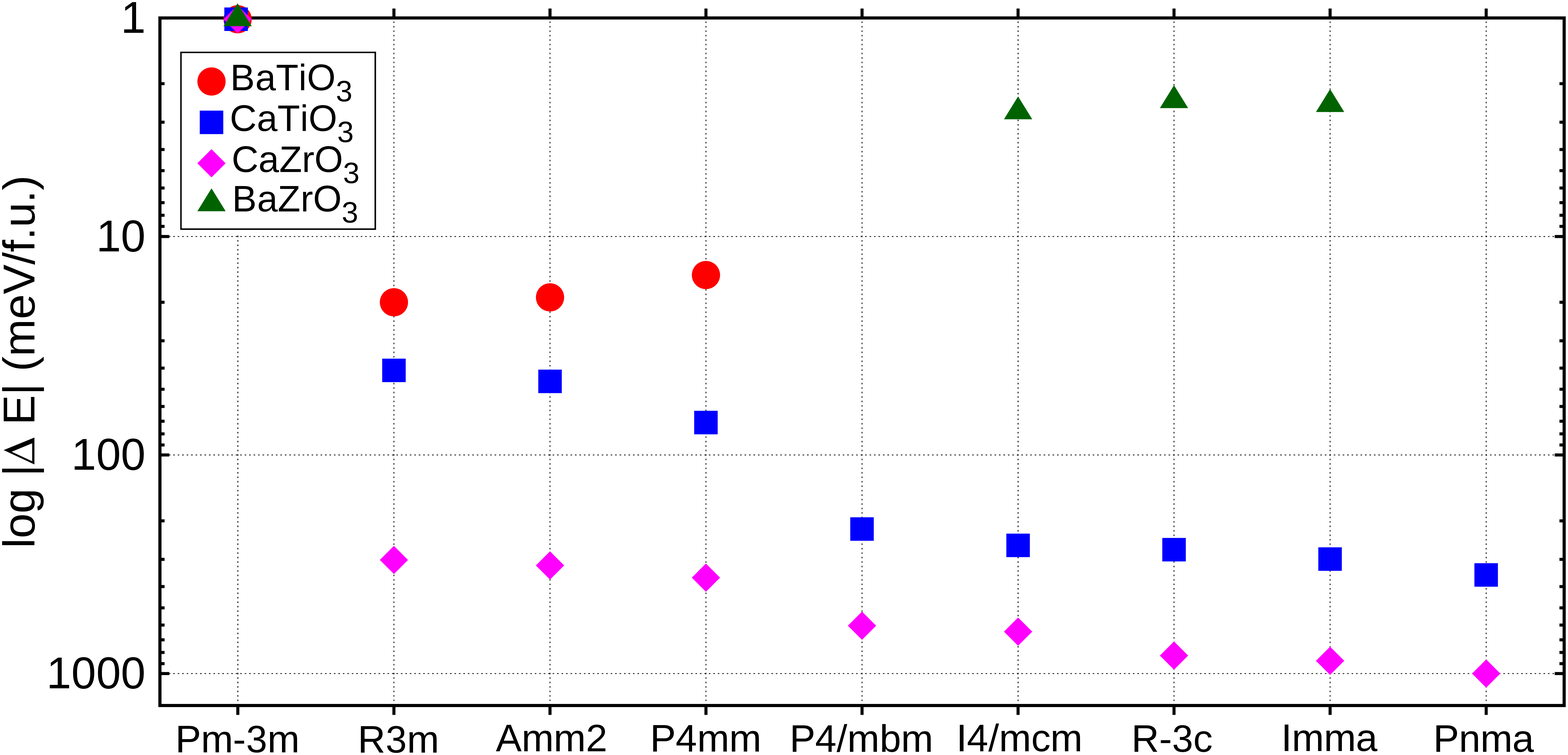}
\caption{\footnotesize Energy landscape (in log scale) for BaTiO$_3$ (red closed circles), CaTiO$_3$ (blue closed squares), CaZrO$_3$ (pink closed rhombus) and BaZrO$_3$ (green closed triangle).
For each $ABO_3$ perovskite, we report the energy gain with respect to the corresponding simple cubic phase, for different optimized phases allowed 
by dynamical properties (phonon spectra Fig.~\ref{bulk_phon}). Polar distortions, characterizing ferroelectric structures, and 
oxygens rotation and/or anti-polar motion, characterizing non-polar distorted structures, have been taken into account separately.
All the structures have been fully relaxed within GGA-WC functional. Main features for the different structures are reported in Table~\ref{modes_bulk}.  }
\label{bulk_phases}
\end{figure}
\begin{table}[h]
\centering
\begin{tabular}{|c|c|c|}
\hline
\multicolumn{3}{|c|}{\bf POLAR STRUCTURES} \\
\hline
Space-group & Notation & Modes \\
\hline
$P4mm$ (99) & a$_0$a$_0$a$_1$ & $\Gamma_4^-$ \\
$Amm2$ (38) & a$_0$a$_1$a$_1$ & $\Gamma_4^-$ \\
$R3m$  (160)& a$_1$a$_1$a$_1$ & $\Gamma_4^-$ \\
\hline
\hline
\multicolumn{3}{|c|}{\bf ANTI-FERRODISTORTIVE} \\
\multicolumn{3}{|c|}{\bf STRUCTURES} \\
\hline
Space-group & Notation & Modes \\
\hline
$P4/mbm$ (127) & a$^0$a$^0$c$^+$ & M$_3^+$ \\
$I4/mcm$ (140) & a$^0$a$^0$c$^-$ & R$_4^+$ \\
$Imma$ (74) & a$^0$b$^-$b$^-$ & R$_4^+$ \\
$R\bar3c$ (167) & a$^-$a$^-$a$^-$ & R$_4^+$ \\
$Pnma$ (62) & a$^-$b$^+$a$^-$ & R$_4^+$, M$_3^+$, X$_5^+$ \\
\hline
\end{tabular}
\caption{\footnotesize Space-group and main active modes in the two allowed classes of distortions. By referring to the simple cubic Brillouin zone:
polar motion is due to instabilities at $\Gamma$~(0,0,0), 
oxygen tilting to instabilities at $M$~($\frac12,\frac12,0$) and $R$~($\frac12,\frac12,\frac12$), and anti-polar motion to instabilities at $X$~($\frac12,0,0$) or to the 
trilinear coupling with the two latter instabilities\cite{benedek1, SrRu_philippe}, as in CaTiO$_3$. The polarization direction associated to the ferroelectric structures and 
Glazer notation associated to the antiferrodistortive structures are also reported.}
\label{modes_bulk}
\end{table}

In the previous section we have seen that the parent compounds can develop two types of instability: a FE instability associated to a polar mode at $\Gamma$ and 
AFD instabilities associated to unstable modes at $M$ and $R$.
Now we quantify the energy difference associated to the condensation of the corresponding (polar and non-polar) distortions with respect to the cubic phase.
The results are shown in Fig.~\ref{bulk_phases}.
A careful crystallographic explanation of the symmetry changes associated to these distortions can be found in Ref.\cite{benedek1}.

\subsubsection{BaTiO$_3$}
In BaTiO$_3$ the energy sequence of the polar phases arising from the condensation of the polar mode is in line with the experimental sequence of phase transitions when lowering 
the temperature ($Pm\bar{3}m \to P4mm \to Amm2 \to R3m$). At the same time, we do not find any other distorted structure as a possible metastable phase. 
To be noticed is that the energy landscape is relatively flat. The $R3m$ ground-state is only $20~meV$/f.u. below the paraelectric $Pm\bar{3}m$ phase and 
the energetic difference between the polar phases is $\simeq 5~meV$/f.u.

\subsubsection{CaTiO$_3$}
Interestingly, in CaTiO$_3$ the energy sequence of the polar phases, arising from the condensation of the unstable polar mode, 
is reversed. In addition, the relative energy difference is larger than in BaTiO$_3$.
In fact, the tetragonal $P4mm$ structure is the lowest energy configuration among the polar ones with a relative energy gain of about $70~meV$/f.u.  
Moreover, additional non-polar structures arising from the condensation of the unstable AFD modes at $M$ and $R$ 
appear as metastable or eventually stable. 
As usual in perovskites with $t<1$ \cite{benedek1}, in fact, the ground-state corresponds to the $Pnma$ ($a^-b^+a^-$) structure which is $\simeq -350~meV$ per f.u. below the  reference cubic phase. 
One-tilt structures, such as $P4/mbm$ ($a^0a^0c^+$) or $I4/mcm$ ($a^0a^0c^-$), 
and the two-tilts $Imma$ ($a^0b^-b^-$) are intermediate metastable structures that appear $\simeq 100~meV$ and $\simeq 50~meV$ per f.u.~above the ground-state, respectively.

\subsubsection{CaZrO$_3$}
In CaZrO$_3$ the energy sequence of both the polar and AFD phases is the same as CaTiO$_3$, but they are shifted down in energy
becoming very spread. The AFD-$Pnma$ ground-state is lower by about $-1~eV$/f.u. than the $Pm\bar3m$ phase. 
This is in tune with the very high transition temperature experimentally observed for the sequence $Pm\bar{3}m \to Pnma$, even if the phonon frequencies related to the unstable modes are 
close to the ones in CaTiO$_3$. 

\subsubsection{BaZrO$_3$}
For BaZrO$_3$ 
we found an interesting energy competion between the antiferroditortive $R\bar3c$ ($a^-a^-a^-$), $Imma$ ($a^0b^-b^-$) and $I4/mcm$ ($a^0a^0c^-$) structures. 
According to the very tiny value of the instability (Fig.~\ref{bulk_phon}(c)), 
the condensation of the oxygens rotations provides an energy gain relative to the cubic phase of about $\simeq 2.5~meV$ per f.u., 
while the three distorted phases have an energy that differs by less than $0.1~meV$/f.u. 
In spite of the negligible energy gain, the amplitude of the AFD distortion is significant. 
The biggest distortion is provided by the tetragonal phase with an angle of rotation of $\sim 4^{\circ}$ about the [001] direction. \\ 

\subsubsection{Ferroelectric phases \& polar modes }
In order to interlock the optimized polar structures with the lattice dynamics, we evaluated the contribution 
of each polar mode to the condensed distortion. 
The overlap matrix is reported in Table~\ref{overlap_bulk}. It is interesting to notice, that the three ferroelectric states are mostly due to the condensation of the unstable optical mode
 for all the three perovskites BaTiO$_3$, CaTiO$_3$ and CaZrO$_3$, so that it is possible to establish a nearly one-to-one correspondence with the 
pattern of distortion associated to the unstable $TO_1$ mode and the displacements as obtained from the structural optimization,
 while the contribution of the $TO_2$ and $TO_3$ modes remains very small.

An important remark is the huge difference in the total distortion $\bm\tau$ between BaTiO$_3$ and the Ca-based perovskites, that allow for possible bigger spontaneous polarization in the latter 
compounds even if the $A$-cation at play is not either stereochemically active or involved in the hybridization with the oxygens, like Ti in BaTiO$_3$. \\

\begin{table}[t]
\centering
\footnotesize
\begin{tabular}{|c|c|c|c|}
\hline
\multicolumn{4}{|c|}{\bf BaTiO$_3$} \\
\hline
mode         & $P4mm$ & $Amm2$ & $R3m$ \\
\hline
${|\bm\tau|}$ & 0.188 & 0.209 & 0.214 \\
\hline
$F_{1u}(TO_1)$ & 0.993 & 0.975 & 0.971 \\
$F_{1u}(TO_2)$ & 0.110 & 0.087 & 0.129 \\
$F_{1u}(TO_3)$ & 0.032 & 0.194 & 0.228 \\
\hline
\multicolumn{4}{|c|}{\bf CaTiO$_3$} \\
\hline
mode         & $P4mm$ & $Amm2$ & $R3m$ \\
\hline
${|\bm\tau|}$ & 0.601 & 0.478 & 0.435 \\
\hline
$F_{1u}(TO_1)$ & 0.985 & 0.970 & 0.970 \\
$F_{1u}(TO_2)$ & -0.171  & -0.203 & -0.199 \\
$F_{1u}(TO_3)$ & 0.033  & 0.129 & -0.143 \\
\hline
\multicolumn{4}{|c|}{\bf CaZrO$_3$} \\
\hline
mode         & $P4mm$ & $Amm2$ & $R3m$ \\
\hline
${|\bm\tau|}$ & 0.903 & 0.886 & 0.868 \\
\hline
$F_{1u}(TO_1)$ & 0.976 & 0.910 & 0.897 \\
$F_{1u}(TO_2)$ & -0.060 & 0.202 & -0.208 \\
$F_{1u}(TO_3)$ & -0.067 & -0.336 & -0.197 \\
\hline
\end{tabular}
\caption{\footnotesize Overlap matrix between the total distorsion $\bm\tau$ of the three optimized polar structures and the eigendisplacements $\bm{\eta}_i$
associated to the $F_{1u}(TO)$ modes of the optimized cubic phase, $\langle\bm{\eta}_i|M|\bm\tau\rangle=\alpha_i$.
The distortion $\bm\tau$ has been normalized such that $\langle\bm\tau|M|\bm\tau\rangle=1$, with $M$ in atomic mass units,
 and results to be defined as ${\bm\tau}=\sum_{i=1,2,3}\alpha_i\bm{\eta}_i$, with $|\bm\tau|=\sqrt{\langle\bm\tau|M|\bm\tau\rangle}$.
Since BaZrO$_3$ has no polar instabilities we reported results only for BaTiO$_3$, CaTiO$_3$ and CaZrO$_3$.}
\label{overlap_bulk}
\end{table}

The previous analysis of the dynamics and energetics associated to the four parent compounds 
has emphasized some similarities and differences. This is helpful for a better understanding 
of properties arising while mixing cations at the $A$- and $B$-sites in (Ba,Ca)TiO$_3$ and Ba(Ti,Zr)O$_3$ solid solutions, respectively.
We will now present results coming from a systematic characterization of the latter systems by also testing and comparing two different approaches:
 the ``virtual crystal approximation'' (VCA) and supercell-based calculations.
 
We note that the parent BaTiO$_3$ and CaTiO$_3$ compounds display inverted sequence of polar phases. Then, for the (Ba,Ca)TiO$_3$ solid solution, 
it can be expected the emergence of a region with strong competition between these phases and a crossing point in energetics. This will be confirmed in the next Section.

\section{Solid solutions: (Ba,Ca)TiO$_3$}
\label{bct_solid}

In the last few years, Ba$_{1-x}$Ca$_{x}$TiO$_3$ (BCT) has started to arouse curiosity in the experimental community as the Ca off-centering seems to play an important role in stabilizing 
ferroelectricity against chemical pressure effects \cite{BCT_exp1,BCT_exp2,BCT_exp3}. 
In particular, in Ref.\cite{BCT_exp2}, Fu and Itoh have characterized the single crystals of BCT in a temperature range from 2~K to 400~K and for compositions ranging from 
$x=0.00$ up to $x=0.34$. They found that the Curie point 
is nearly independent of the Ca-concentration for the $Pm\bar{3}m \to P4mm$ transition, whereas there is a shift of the $P4mm \to Amm2$ and $Amm2 \to R3m$ phase transitions 
toward lower temperatures. Accordingly, the consequence of Ca-substitution is the stabilization of the tetragonal ferroelectric phase. 

Let us now analyze the dynamics and energetics as predicted by means of first-principles calculations. We first report results from the VCA approach in Sec.~\ref{bct_vca}, 
then from the use of supercells in Sec.~\ref{bct_super}.

\subsection{Virtual Crystal Approximation (VCA)}
\label{bct_vca}

\subsubsection{Lattice parameter}

First, we report the evolution of the lattice parameter of the $Pm\bar{3}m$-cubic phase as obtained from the structural optimization within VCA. 
Because of the reduced volume of CaTiO$_3$ with respect to BaTiO$_3$ (Table~\ref{eigen_bulk}), the lattice parameter decreases monotonically with the Ca-concentration, 
but the trend deviates from the linearity of the Vegard's law, as shown in Fig.~\ref{acell_vca_bct}(a). 

As for the pure compounds, for solid solutions we can define the tolerance factor as $t= (\overline{r_A}+r_O)/\sqrt2(r_B+r_O)$, with $\overline{r_A}=(1-x)r_{Ba}+(x)r_{Ca}$. 
Values of the ionic radii, $r_i$, for the pure atoms have been taken from Ref.\cite{ionic_radii1}. For increasing $x$, $t$ decreases from $1.06$ to $0.97$ 
reaching 1 at $x=0.6$, as reported on top of Fig.~\ref{acell_vca_bct}.

\subsubsection{Dynamical properties}

\begin{figure}[t]
\centering
\includegraphics[width=\columnwidth]{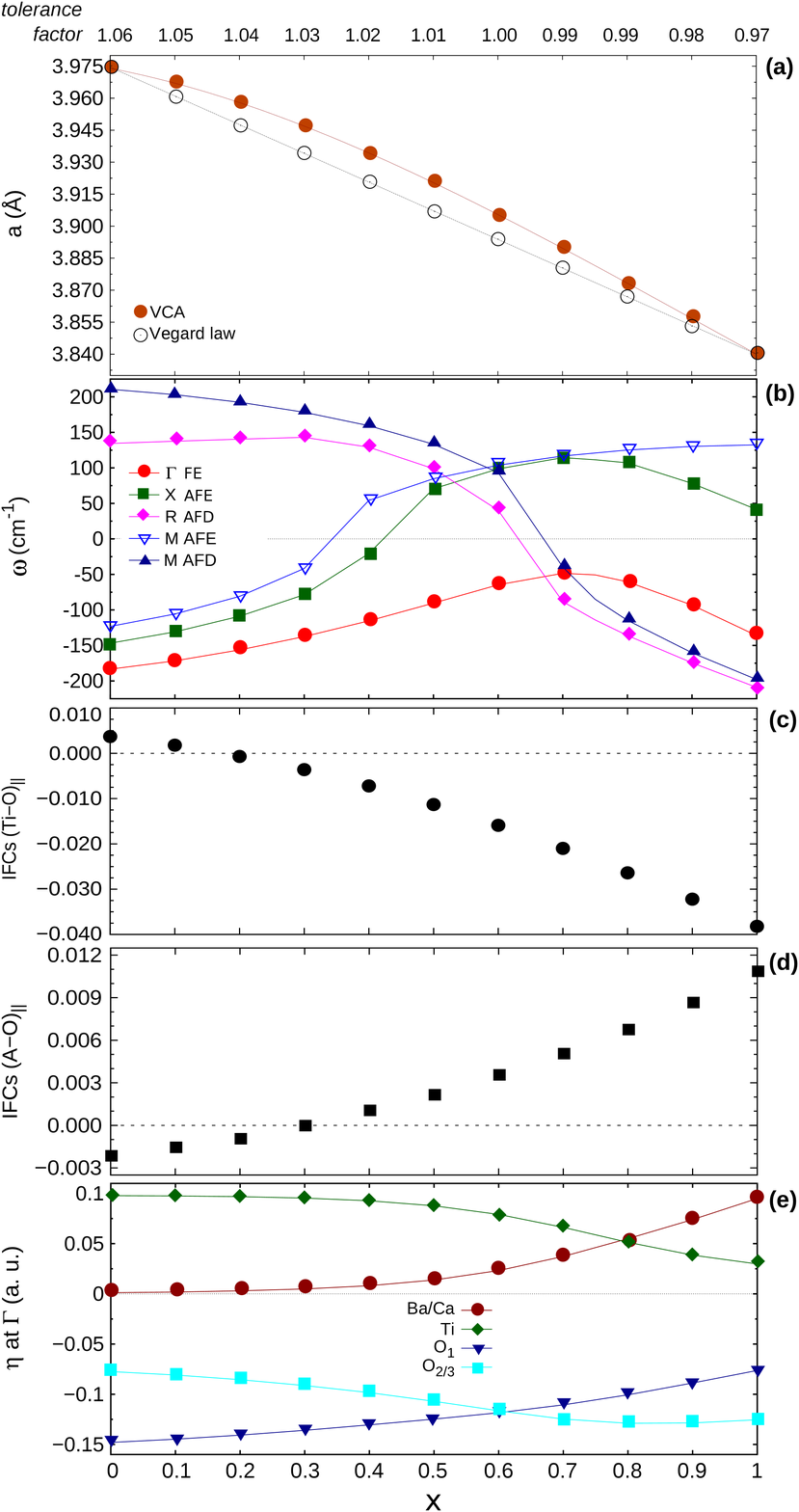}
\caption{\footnotesize Evolution of different lattice properties as a function of $x$-composition in Ba$_{1-x}$Ca$_{x}$TiO$_3$ as obtained by the use of VCA. 
(a) Cubic lattice parameter $a_{cell}$ (in \AA). Vegard's law has been built on the theoretical values of the cubic BaTiO$_3$ and CaTiO$_3$ reported in Table~\ref{eigen_bulk}: 
$a(x)=(1-x)a_{BTO}+(x)a_{CTO}$. 
(b) Trend of the lowest phonon frequencies at the high-symmetry points of the simple cubic Brillouin zone. 
Evolution of the total Ti-O (c) and $A$-O (d) Interatomic Force Constants (IFCs in Ha/Bohr$^2$).
(c) Evolution of normalized eigendisplacements (in $a.u.$) associated to the unstable polar mode at the $\Gamma$-point of the simple cubic Brillouin zone. 
Lines are guide for the eyes.}
\label{acell_vca_bct}
\end{figure}

Then, we analyze the evolution of the lowest phonon frequencies at the high symmetry points of the Brillouin zone of the cubic reference. 
DFPT calculations have been performed on the optimized structures.
We find that, for increasing Ca-concentration, the original polar instabilities of BaTiO$_3$ 
changes character by becoming confined to the center of the Brillouin zone (AFE instabilities at $X$ and $M$ disappear), while the AFD modes
become unstable, as shown Fig.~\ref{acell_vca_bct}(b). Changes in the phonon instabilities are linked to the evolution of $t$. 
In fact, the polar instability dominates until $t$ becomes smaller than one and the AFD modes become largely unstable. 
Specifically, 
the lattice dynamics is BaTiO$_3$-like for $0 \leq x \leq 0.2$. The unstable polar mode at $\Gamma$ is mainly sustained by the destabilizing 
Ti-O interaction, while the $A$-O interaction stays mostly repulsive by resulting in the inactivity of the $A$-site, as reproduced in Fig.~\ref{acell_vca_bct}(c,d,e). 
Accordingly, the instabilities at the $X$- and $M$-points are also due to Ti-O polar motion. For $0.3 \leq x \leq 0.6$, the scenario starts to change. 
The polar instability becomes weaker as it results from the important reduction in the amplitude of the associated frequencies. 
Particularly, the polar distortion remains unstable at $\Gamma$, while it becomes progressively stable at $X$ and $M$. 
These smooth changes can be related to smooth changes in the type of interaction between the cations and the oxygens. A change of sign is observed along 
both the Ti-O and $A$-O interaction that corresponds to a strong competition between attractive and repulsive forces in determing the nature and character of the polar instability: 
a reduction of the amplitude of the Ti-motion corresponds to increasing negative value of the Ti-O interaction, while an increasing contribution to the polar distortion of the $A$-site 
arises from positive values of the $A$-O interaction, as shown in Fig.~\ref{acell_vca_bct}(c,d,e). 
Finally, for $0.7 \leq x \leq 1.0$, the lattice dynamics becomes CaTiO$_3$-like. Accordingly, instabilities related to oxygens rotations appear at the $M$- and $R$-points of 
the cubic Brillouin zone with larger associated frequencies than the polar one, that remains unstable only at $\Gamma$. This scenario results from strong destabilizing $A$-O interactions and 
largely repulsive Ti-O interactions. Accordingly, the character of the polar distortion also changes by becoming largely driven by the motion of the $A$-site with respect to the $B$-one.

Moreover, changes in the phonon behaviour also affect the dynamics of the oxygens. 
The increasing contribution to the distortion from the $A$-site, i.e. increasing long-range forces between the $A$-cation and oxygens,
 favors the motion of the planar oxygens with respect to the apical one (labelled 
O$_{2/3}$ and O$_1$ in Fig.~\ref{acell_vca_bct}(e), respectively). Specifically, the contribution is reversed when going from $t>1$ to $t<1$. \\

The latter analysis of the evolution of the dynamical properties when going from the BaTiO$_3$-rich region to the CaTiO$_3$-rich one in the Ba$_{(1-x)}$Ca$_{(x)}$TiO$_3$ ``virtual-system" 
reveals the presence of two parallel mechanisms: the progressive weakening of the long-range forces between Ti and O atoms in favor of their strengthing between the $A$-site and oxygens. 
In terms of character of the phonons instabilities, this change corresponds to a smooth evolution from the $B$-driven into the $A$-driven distortions
associated to the parents BaTiO$_3$ and CaTiO$_3$, respectively. 
The reasons behind these two phenomena are also different and complementary. In fact, Ca-doping of the virtual $A$-site produces: 
(i) varying interatomic force constants between the $A$-O atomic pair in favors of destabilizing long-range forces sustained by the progressive lowering of the $A$-cation stiffness 
when going from BaTiO$_3$ to CaTiO$_3$ (see Table~\ref{sfc_bulk}); 
(ii) reduction of the volume, that can be considered as increasing isotropic pressure on the $A$TiO$_3$ system and, therefore, shortnening of the Ti-O bond lengths. 
The latter effect produces increasing stiffness of Ti-atoms, that results in the change of sign of the Ti-O interatomic force constants and associated weakening of the $B$-type ferroelectricity.
In fact, as investigated at the first-principle level in Refs.~\cite{cohen_BTO, phil_europhys, eric_BTO} and as it turns out when comparing graphs (a) and (c) reported in Fig.~\ref{acell_vca_bct}, 
the balance between short-range and long-range forces between the Ti and O atoms is strongly sensitive to pressure. 
At variant, the varying composition does not affect significantly the $A$-$B$ interaction, that in fact remains almost the same as in the pure parent compounds 
(see Tables~\ref{ifc_bulk_local} and \ref{ifc_BCT_12_super}).

\subsubsection{Energetics landscape}

\begin{figure}[h]
\centering
\includegraphics[width=\columnwidth]{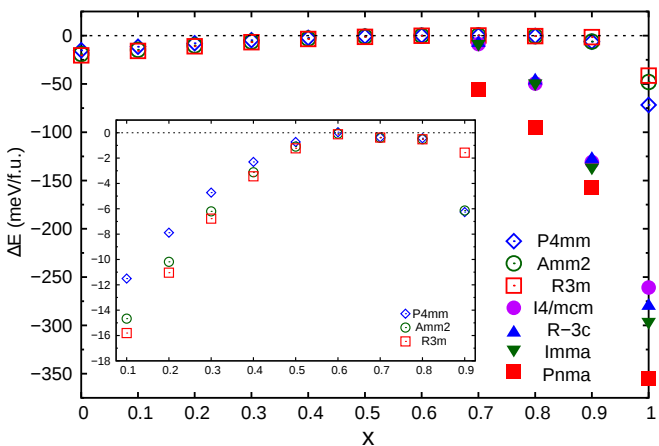}
\caption{\footnotesize Energy gain with respect to the simple cubic phase for different optimized structures allowed by dynamical properties within VCA as a function 
of $x$-composition in Ba$_{1-x}$Ca$_{x}$TiO$_3$. 
Structures properties reported in Table~\ref{modes_bulk}. The inset shows a zoom on the energetics of the polar phases for $0.1 \leq x \leq 0.9$.}
\label{bct_vca_energy}
\end{figure}

As we previously did for the pure parent compounds, now that the main instabilities of BCT ``virtual-system" have been identified,
 we look at the energy competition between different phases arising from the condensation of the corresponding modes in order to obtain an overview of the energetics 
trend as a function of varying concentration. 
As shown in Fig.~\ref{bct_vca_energy}, the main effect of the Ca-substitution 
is to reduce the energy gain both between the three ferroelectric phases and with respect to the cubic phase by making them strongly competitive, while 
additional antiferrodistortive structures appear as metastable in the CaTiO$_3$-rich region by providing a non-polar ground-state.

Such behaviour proceeds from the evolution of the dynamical properties previously analyzed. In fact, an important energy competition between polar phases starts at $x=0.3$, that corresponds 
to the concentration where we observe the change of sign of the Ti-O IFCs (see Fig.~\ref{bct_vca_energy}(c)). 
Then, the energies of the polar phases smoothly converge to that of the cubic one in line with the progressive weakening of the polar instability. 
Noteworthy is that the inversion in the energy sequence of the three polar phases happens at $x=0.9$ where we 
observe the inversion of the dominant character of the polar instability between virtual $A$-cation and Ti (see Fig.~\ref{bct_vca_energy}(e)).
 
Additional structures with oxygens rotations appear with an higher energy gain from $x=0.7$ proceeding from the appearence of the AFD phonons instabilities 
(see Figs.~\ref{acell_vca_bct}(b) and~\ref{bct_vca_energy}). 
It is also noteworthy that the contribution to the polar distortion arising from the apical and planar oxygens is inverted from this point, 
with the dominant motions of the latter ones (see Fig.~\ref{acell_vca_bct}(e)).

\subsubsection{Polarization and piezoelectric response}

\begin{figure}[t]
\centering
\includegraphics[width=\columnwidth]{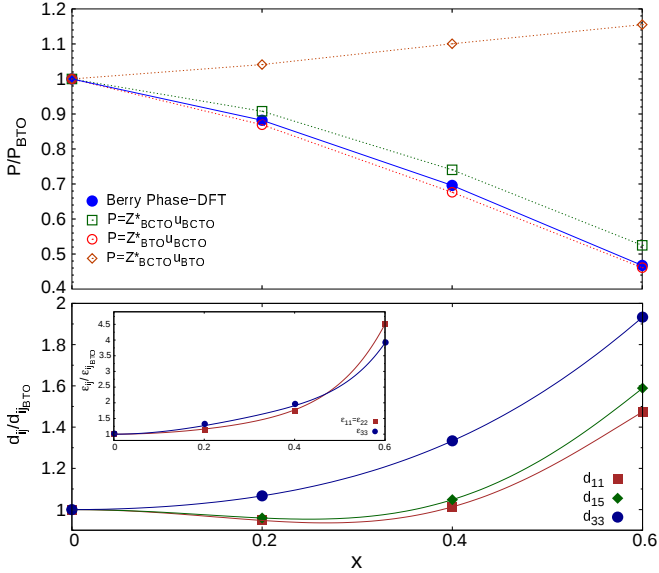}
\caption{\footnotesize Evolution of polarization, piezoelectric and free-stress dielectric coefficients in the rhombohedral-$R3m$ phase
as a function of $x$-composition in Ba$_{1-x}$Ca$_{x}$TiO$_3$ within VCA.
Values are normalized to the ones of $R3m$-BaTiO$_3$: $P_s^{BTO}\simeq 38~\mu C/cm^2$; $d_{11}^{BTO}\simeq 76~ pC/N$, $d_{15}^{BTO}\simeq 270~ pC/N$ and $d_{33}^{BTO}\simeq 15~ pC/N$; 
$\varepsilon_{11}^{BTO}\simeq 208$ and $\varepsilon_{33}^{BTO}\simeq 11$. 
Upper panel: variation of the polarizaton as directly computed by DFT calculations (closed circles) 
and by Born effective charges approximation to disentangle the contribution of varying atomic charges and displacements. Lower panel: trend of the $d_{ij}$ coefficient and 
free-stress dielectric response $\varepsilon_{ij}$.}
\label{bct_vca_pol}
\end{figure}

Proceeding from the previous observation that for $0.0 \leq x \leq 0.6$ the BCT ``virtual-system" experiences 
a progressive weakening of the ferroelectric instability 
and that no other phonons are unstable, we now evaluate the evolution of the spontaneous polariation, $P_s$, and piezoelectric coefficients, $d_{ij}$. 
According to the VCA results,
we perfomed calculations on the lowest energy rhombohedral-$R3m$ structure (Fig~\ref{bct_vca_energy}). Values are normalized to the calculated ones of pure $R3m$-BaTiO$_3$, 
that are $P_s^{BTO}\simeq 38~\mu C/cm^2$ and $d_{11}^{BTO}\simeq 76~ pC/N$, $d_{15}^{BTO}\simeq 270~ pC/N$ and $d_{33}^{BTO}\simeq 15~ pC/N$. 

Specifically, in Fig.~\ref{bct_vca_pol}, we report $P_s$ as obtained both via first-principles calculations by means 
of the Berry Phase Theory \cite{berry_phase} (blue closed circles) and via the Born effective charges by means of the approximation 
$P_{s,\alpha}=\frac{1}{\Omega}\sum_{k,\beta}Z^{*}_{k,\alpha\beta}\Delta\tau_{k,\beta}$ \cite{phil_Z} (green open squares).
It results that $P_s$ decreases as function of increasing Ca-composition in tune with the decreasing energy gain associated to the polar phases in this range of composition, 
i.e. the weakening of the polar instability detected within VCA (Figs.~\ref{acell_vca_bct} and \ref{bct_vca_energy}). 
In order to clarify if changes in the polarization are mostly due to varying effective charges or atomic distortions, we disentangled the two contributions.
In the first case, we kept constant the atomic displacements of pure $R3m$-BaTiO$_3$ and took into account the different effective charges associated to each BCT compositions (orange open rhombus). 
In the second case, we kept constant the Born effective charges of pure $R3m$-BaTiO$_3$ and considered the distortion arising from different compositions (red open circles). 
It results that the evolution of $P_s$ has to be widely ascribed to decreasing atomic distortion in BCT ``virtual-system" in such range of composition, as shown in Fig.~\ref{bct_vca_pol} (top). 
In fact, this behaviour is in line with the evolution of the eigendisplacements associated to the unstable polar mode at $\Gamma$ 
with a pronounced reduction of the titanium and oxygens motion Fig.~\ref{acell_vca_bct}(e).

Despite the decrease of polarization, the piezoelectric response increases with the Ca-concentration (Fig.~\ref{bct_vca_pol} bottom). 
This trend is due to the fact that the free-stress dielectric response, $\varepsilon_{ij}$, remarkably increases as well, as shown in the inset of Fig.~\ref{bct_vca_pol}. 
By writing down the piezoelectric coefficient as $d_{ik}=\frac{de_{s,i}}{dE_k}\propto(\chi_{j,k} P_{s,l}+P_{s,j}\chi_{l,k})$ \cite{piezo_review}, it is easier to understand 
such behaviour: near a phase transition, where the lowest-frequency polar mode, i.e. the soft mode, goes to zero, the dielectric response diverges \cite{rabe_philippe}. 
Accordingly, the calculated values of the lowest-frequency polar modes in the $R3m$-phase evolve like 169, 160, 124, 73 (in $cm^{-1}$)
for 0.0, 0.2, 0.4 and 0.6 $x$-composition, respectively.
Therefore, the softening of the polar mode overcomes the reduction of $P_s$ in the piezoelectric response within the VCA approach.\\ 

\subsection{Supercell Calculation}
\label{bct_super}

In order to check the validity of VCA and to better characterize the impact of the (Ba,Ca) substitution on the dynamics of the system
we performed direct DFT calculations on the Ba$_{0.875}$Ca$_{0.125}$TiO$_3$ and Ba$_{0.50}$Ca$_{0.50}$TiO$_3$ compositions by means of supercells.
Details about the atomic arrangements have been provided in Sec.~\ref{supercell_description}.

\subsubsection{Ba$_{0.875}$Ca$_{0.125}$TiO$_3$}

\begin{table*}[t]
\centering
\footnotesize
\begin{tabular}{|c||c|c|c|c||cc|c|c|c|}
\hline
\multicolumn{10}{|c|} {\bf Ba$_{0.875}$Ca$_{0.125}$TiO$_3$} \\
\hline
 \multicolumn{5}{|c||} {\bf VCA} & \multicolumn{5}{c|} {\bf 2x2x2 CELL} \\
\hline
Atoms & Direction &  Total &   DD    &   SR &  Atoms &                         &   Total &   DD    &   SR  \\
\hline
$A-A'$ & & &  &                                      & Ba-Ba'  &($\parallel$) & -0.012 & -0.006 & -0.006 \\
       & ($\parallel$) & -0.011 & -0.005 & -0.006   & d=3.959 \AA   & ($\perp$)    & +0.004 & +0.003 & +0.001 \\[4pt]
       & ($\perp$)     & +0.004 & +0.003 & +0.001   & Ca-Ba &($\parallel$) & -0.009 & -0.005 & -0.004  \\
       & d=3.967 \AA & &   &                          &  d=3.959 \AA  & ($\perp$)    & +0.004 & +0.003 & +0.001  \\
\hline
$B-B'$ & ($\parallel$) & -0.070 & -0.038 & -0.032     & Ti-Ti'  & ($xx'$) & -0.072 & -0.040 & -0.033 \\
       & ($\perp$) &     +0.008 & +0.019 & -0.011    & d=3.929 \AA & ($yy'$) & +0.008 & +0.020 & -0.012 \\
       & d=3.967 \AA &   &  &  &                                  & ($zz'$) & +0.009 & +0.020 & -0.011 \\
\hline
$A-B$ & & &      &                                  & Ba-Ti   &($\parallel$) & -0.030 & -0.022 & -0.008 \\
      & ($\parallel$) & -0.030 & -0.022 & -0.008   & d=3.437 \AA   & ($\perp$)    & +0.014 & +0.011 & +0.003 \\[4pt]
      & ($\perp$) &     +0.014 & +0.011 & +0.003   & Ca-Ti     &($\parallel$) & -0.023 & -0.022 & -0.001 \\
      & d=3.436 \AA & &       &                      & d=3.403 \AA   & ($\perp$)    & +0.013 & +0.011 & +0.003 \\
\hline
$B-O_1$    & & &      &                                 & Ti-O$_1$  & ($\parallel$) & +0.005 & +0.236 & -0.231 \\
          & ($\parallel$) & +0.002 & +0.242 & -0.241    & d=1.994 \AA  & ($\perp$) & -0.020 & -0.045 & +0.025 \\[4pt]
          & ($\perp$) &     -0.020 & -0.045 & +0.025   &  Ti-O$_2$  & ($\parallel$) & -0.007 & +0.253 & -0.260 \\
          &  d=1.984 \AA &           &    &           &  d=1.965 \AA  & ($\perp$) & -0.020 & -0.044 & +0.024   \\
\hline
$A-O_1$   & & &     &                                 & Ba'-O$_1$  & ($xx'$) & -0.004 & +0.012 & -0.016 \\
          & & &      &                                & d=2.783 \AA  & ($yy'$) & -0.011 & -0.016 & +0.005 \\
          & & &       &                               &               & ($zz'$) & -0.004 & -0.006 & +0.002 \\[4pt]
          & ($xx'$) & -0.001 & +0.012 & -0.013        & Ba'-O$_2$  & ($xx'$) & -0.003 & +0.012 & -0.015 \\
          & ($yy'$) & -0.004 & -0.006 & +0.002          & d=2.800 \AA    & ($yy'$) & -0.005 & -0.006 & +0.001 \\
          & ($zz'$) & -0.011 & -0.016 & +0.005             &                 & ($zz'$) & -0.011 & -0.016 & +0.005 \\[4pt]
          & d=2.805 \AA & &         &                         & Ca-O$_2$    & ($xx'$) & +0.011 & +0.011 & -0.000  \\
           & & &        &                        & d=2.750 \AA  & ($yy'$) & -0.005 & -0.006 & +0.001 \\
          & & &         &              &           & ($zz'$) & -0.011 & -0.016 & +0.005 \\
\hline
\end{tabular}
\caption{\footnotesize Interatomic force constants for Ba$_{0.875}$Ca$_{0.125}$TiO$_3$ from  VCA and super cell calculations.
Units are in Ha/Bohr$^2$. Directions $xx'~(\parallel)$, $yy'~(\perp)$ and  $zz'~(\perp)$ refer to local coordinates system of the different pairs of atoms.
Distances (in \AA~) between the selected atoms are also reported. Atoms' notation in the second column is consistent with Fig.~\ref{polar_phases_bct}(a)}
\label{ifc_BCT_12_super}
\end{table*}

Phonons calculations performed at the $\Gamma$-point of the cubic $Pm\bar{3}m$-supercell of Ba$_{0.875}$Ca$_{0.125}$TiO$_3$ (Fig.~\ref{super_12}(a)) have revealed several instabilities 
related to polar modes. The most unstable one, with $\omega\simeq 169i~cm^{-1}$, is associated to a polar distortion driven by all the Ti-atoms and the single Ca-atom 
against the oxygens. Ba-atoms are almost fixed as in the parent BaTiO$_3$. 
Therefore, Ca is polar active on the $A$-site already at low concentration.
The different dynamics can be analyzed in terms of interatomic force constants between the $A$-cations and oxygens in the investigated solid solution.
In fact, the $A$-O interaction results to be opposite if we focus on Ba or Ca atoms in line with the dynamical properties of the respective pure compounds (Table~\ref{ifc_bulk_local}). 
As reported in Table~\ref{ifc_BCT_12_super}, the Ba-O interaction remains dominated by repulsive forces, while the Ca-O interaction is largely dominated by the destabilizing long-range interaction. 
Moreover, by looking at the atomic pair properties reported in Table~\ref{ifc_BCT_12_super}, a further effect of the Ca-presence is to make oxygens 
inequivalent even in the high centro-symmetric reference. This effect is due to an internal 
structural relaxation around the Ca-atom because of its smaller size than Ba. This mechanism strongly affects the Ti-O interaction. In fact, the Ti atom experiences 
shorter and longer bonds with the corresponding apical oxygen, that produce strong competition between long- and short-range forces: along the longer bond the dipole-dipole interaction 
overcomes the repulsion, while it is the opposite along the shorter bond (see values for the supercell case in Table~\ref{ifc_BCT_12_super}). 
For this specific BaTiO$_3$-rich concentration, such delicate balance 
between the attractive and repulsive interactions along the Ti-O bond results in the weakening of the $B$-site driven ferroelectricity.

The intriguing manifestation of this phenomenon is the achievement of a quasi-degeneracy between different polar states.
In fact, we analyzed five different polar states corresponding to the tetragonal, orthorhombic, rhombohedral, monoclinic and triclinic phases with $P4mm$, $Amm2$, $R3m$, $Pm$ and $P1$ 
space-group symmetry, respectively. 
Specifically, the everage gain in energy with respect to the $Pm\bar{3}m$-cubic phase is about $-20$ meV/f.u. as for the parent BaTiO$_3$. Nevertheless, the maximum energy difference between 
the different polar phases is of about $0.5$ meV/f.u.

According to the latter isotropic energy landscape, the spontaneous polarization results to be essentially the same for all these ferroelectric phases, that is 
$P_s\simeq 39.5~\mu C/cm^2$ as obtained via Berry phase calculations. This value is slightly higher than $P_s$ of pure BaTiO$_3$ in the $R3m$-phase showed in Fig.~\ref{bct_vca_pol} of 
the previous section. 
Moreover, we also calculated the $d_{ij}$ coefficients of the piezoelectric tensor in the stable $R3m$-phase. The $d_{33}$ component, parallel to the polarization direction, remains 
unchanged with respect to BaTiO$_3$, while the $d_{11}$ and $d_{15}$ components, tranversal to the polar axis, are considerably enhanced. 
Specifically they are: $d_{33}\simeq 15$, $d_{11}\simeq 344$ and $d_{15}\simeq 1458$ in $pC/N$. 

\begin{figure}[b]
\centering
\includegraphics[width=8cm]{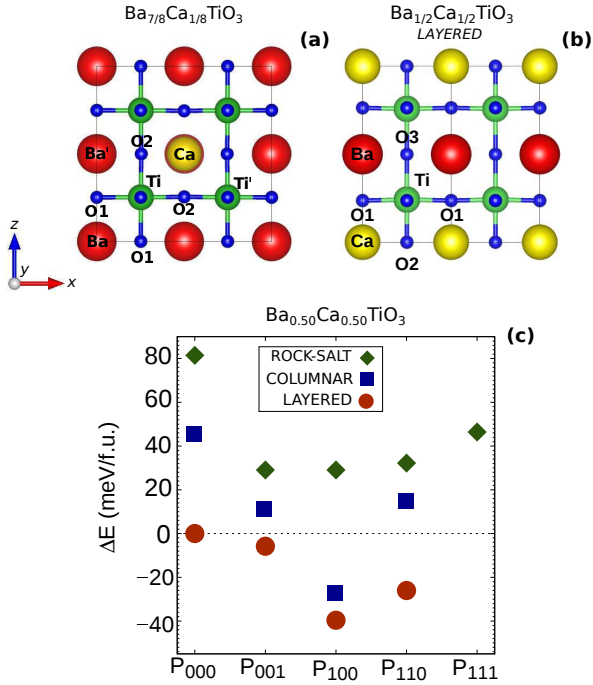}
\caption{\footnotesize Schematic representation of the atomic arrangement in Ba$_{0.875}$Ca$_{0.125}$TiO$_3$ (a) and in layered-Ba$_{0.5}$Ca$_{0.5}$TiO$_3$(b). 
Direction along $y$ is equivalent to the $x$-one. Labels help the visualization of the main atomic pair analyzed in Tables~\ref{ifc_BCT_50} and~\ref{ifc_BCT_12_super}.
(c): Energy gain (in meV/f.u.) with respect to the highy-symmetry $P4mmm$-layered structure for different optimized polar states
 in the Ba$_{0.5}$Ca$_{0.5}$TiO$_3$ supercells. The $x$-axis notation refers to the direction of the components of polarization. }
\label{polar_phases_bct}
\end{figure}

\subsubsection{Ba$_{0.5}$Ca$_{0.5}$TiO$_3$}

First result in Ba$_{0.5}$Ca$_{0.5}$TiO$_3$ ((Fig.~\ref{super_12}(b-d))) is the preference of an ordered configuration composed by alternating layers of the same type of $A$-cations 
with respect to a columnar and rocksalt ordering. 
The relative energy gain of the high-symmetry columnar and layered configurations with respect to the rocksalt one is about $-35.5~meV/$f.u. and $-81.4~meV/$f.u., respectively.

\begin{table}[t]
\centering
\footnotesize
\begin{tabular}{|c||c|c||cc|c|}
\hline
\multicolumn{6}{|c|} {\bf Ba$_{0.50}$Ca$_{0.50}$TiO$_3$} \\
\hline
 \multicolumn{3}{|c||} {\bf VCA} & \multicolumn{3}{c|} {\bf 1x1x2 CELL} \\
\hline
Atoms & Direction &  Total &   Atoms &  &   Total  \\
\hline
$B-O_1$   &               &           &  Ti-O$_1$  & ($\parallel$) & -0.009 \\
          &            &      & d=1.956 \AA  & ($\perp$) & -0.019  \\[4pt]
          & ($\parallel$) &     -0.011  &  Ti-O$_2$  & ($\parallel$) & -0.047 \\
          &  ($\perp$) &     -0.020   &  d=1.890 \AA  & ($\perp$) & -0.017  \\[4pt]
          & d=1.960 \AA &              &  Ti-O$_3$  & ($\parallel$) & +0.010 \\
          &              &               &  d=2.011 \AA  & ($\perp$) & -0.019  \\
\hline
$A-O_1$   & &                                  & Ba-O$_1$  & ($xx'$) & -0.001 \\
          & &                                  & d=2.837 \AA  & ($yy'$) & -0.005 \\
          & &                                &               & ($zz'$) & -0.011 \\[4pt]
          &  &                            & Ba-O$_3$  & ($xx'$) & -0.005 \\
          & ($xx'$) & +0.002                & d=2.765 \AA    & ($yy'$) & -0.005 \\
          & ($yy'$) & -0.005                &     & ($zz'$) & -0.011 \\[4pt]
          & ($zz'$) & -0.011                 & Ca-O$_{1}$    & ($xx'$) & +0.010  \\
          & d=2.772 \AA &                              & d=2.690 \AA  & ($yy'$) & -0.006 \\
          & &               &                              & ($zz'$) & -0.012 \\[4pt]
          & &                              & Ca-O$_{2}$    & ($xx'$) & +0.011  \\
          & &                             & d=2.765 \AA  & ($yy'$) & -0.005 \\
          & &                              &              & ($zz'$) & -0.012 \\
\hline
\end{tabular}
\caption{\footnotesize Interatomic force constants for Ba$_{0.50}$Ca$_{0.50}$TiO$_3$ within VCA and the layered supercell.
Units are in Ha/Bohr$^2$. Directions $xx'~(\parallel)$, $yy'~(\perp)$ and  $zz'~(\perp)$ refer to local coordinates system of the different pairs of atoms.
Distances (in \AA~) between the selected atoms are also reported. Atoms' notation in the second column is consistent with Fig.~\ref{polar_phases_bct}(b)}
\label{ifc_BCT_50}
\end{table}

Then, as previously, we performed DFPT calculations on the three high-symmetry reference supercells.
All three configurations display several instabilities.
In all of them the most unstable mode corresponds to a polar distortion driven by both the Ti and Ca atoms like in Ba$_{0.875}$Ca$_{0.125}$TiO$_3$, but 
with an average major contribution from calcium. The phonon frequency associated to that mode 
is $\omega \simeq 141i~cm^{-1}$ in all three cases. 
Additionally, AFD modes related to oxygens rotation become unstable. However, the energy gain carried to the system by the condensation of the latter distorions alone is lower 
than the one coming from the polar distortions, that determine the ground-state. 

Accordingly, the most stable polar phase results to be the $P2mm$-structure with the polarization along the [100] direction of the layered atomic arrangement. 
The associated $P_s$ is of about $51~\mu C/cm^2$ and it is considerably greater than the one calculated in $P4mm$-BaTiO$_3$, that is $P_s\simeq 34~\mu C/cm^2$.
Moreover, in order to see eventual effects of the atomic configuration on the energy competition between different polar states, 
we condensed different patterns of polar distortions in the three arrangements. The associated energy landscape is quite spread as reported in Fig.~\ref{polar_phases_bct}(c). 
It results that, for $x=0.50$, the system prefers a state with only ``one-component'' polarization independently of the atomic configuration. 
The $R3m$-like state with polarization along the [111] pseudo-cubic direction is largely penalized.

Also in this case, changes in the dynamical properties (i.e. substantial contribution from Ca-atoms to the polar distortion and the appearance of unstable AFD modes) and in the 
energetics of the polar phases can be analyzed in term of interatomic force constants. As made clear in Tables~\ref{ifc_BCT_50} and~\ref{ifc_BCT_12_super}, 
the main features are the same as the ones described for the $x=0.125$ Ca-concentration. 
However, the higher concentration of calcium produces here: (i) a global weakening of the long-range interaction between Ti and O atoms in favor of repulsive forces via 
further shortened Ti-O bonds lenghts, i.e. further reduction of the [TiO$_6$]-octahedral volume; this can be the reason behind the tendency of the system to prefer the 
``tetragonal-like'' polar state; (ii) the presence of unstable AFD modes because of the reduced number compressed Ba-O distances otherwise preventing the octahedral rotations.

\subsection{VCA vs SUPERCELL calculations}

By comparing results from supercells-based calculations with the ones from VCA,
it seems that the latter approach is successful to provide an average and qualitative identification of the main changes occuring in the (Ba,Ca)TiO$_3$ solid solution. 
However, it cannot provide a proper description of the microscopic mechanism behind properties. 
The reason of such a limitation is that, by construction, VCA considers all $A$ atoms as equivalent and forces them to behave similarly while, in fact, 
they want to adopt different behaviors.
The importance of considering the actual local cations is clear by comparing the quantities reported in Tables~\ref{ifc_BCT_50} and~\ref{ifc_BCT_12_super}, where we clearly see different results are obtained depending on the method. 
In fact, as deeply discussed above, evolutions in the dynamics as well as in the energy landscape are strongly related to the presence of calcium that 
directly interacts with the oxygens and, indirectly, affects the Ti-O interaction via local changes of the structure.
Therefore, the progressive weakening of ferroelectricity, within VCA, is not representative of the real behaviour. 
In fact, the actual weakening of the Ti-driven ferroelectricity due to decreasing volume is progressively compensated by the Ca-driven ferroelectricity. 
Accordingly, the inversion in the sequence of polar phases when going from BaTiO$_3$ to CaTiO$_3$ appears already at $x=0.50$ within the supercell, while it is expected for $x\geq0.9$ within VCA.
Additionally, an opposite trend of the spontaneous polarization is also obtained within the two approaches: decreasing values for larger $x$ within VCA, while increasing ones within the supercells.

As a global result, the success of VCA in detecting the weakening of the $B$-driven ferrolectricity is due to the fact that the Ca-doping mainly induces a steric effect 
because of the volume contraction when the Ca-concentration increases, while the electronic properties are not primary. 
In fact, the top of the valence is mostly characterized 
by the O $p$ and Ti $d$ states entering the hybridization mechanism, while the $s$ and $p$ states of Ba and Ca stay well below the Fermi level. 
Nevertheless, 
the trends and microscopic mechanisms involved as predicted by means of supercells calculations are in line with experimetal observations\cite{BCT_exp2, BCT_levin}. 
This remark confirms that, in order to provide a proper first-principles characterization of (Ba,Ca)TiO$_3$, VCA-based approaches 
can provide some trend, but are not appropriate to explain the underlying physics; for the latter, 
supercells-based calculations taking explicitly into account the different nature of the cations are more suitable. 

\section{Solid solution: Ba(Ti,Zr)O$_3$}
\label{btz_solid}
Ba(Ti,Zr)O$_3$ (BTZ) solid solutions involve the homovalent substitution between Ti$^{4+}$ and Zr$^{4+}$, that are nevertheless quite different atoms both for the ionic radii 
(0.605 \AA~and 0.72~\AA~respectively \cite{ionic_radii1}) and the electronic configuration (Ti 4s$^2$ 3d$^2$ and Zr 4d$^2$ 5s$^2$) entering the hybridization mechanism, that for instance
leads the ferroelectricity in BaTiO$_3$. 

The first experimental investigation of the phase diagram of the BaTiO$_3$-BaZrO$_3$ binary system dates back to the 1956 with the work of Kell and Hellicar\cite{first_BTZ}, 
reporting the abrupt effect of Zr-concentration on the decreasing of the Curie point for the three ferroelectric phases of BaTiO$_3$. Subsequently, 
other solid experimental investigations came out in the late '90s, 
when J. Ravez {\emph et al.} in Refs.\cite{BTZ_exp1,BTZ_exp2} provided a
clear distinction of phases in the BaTi$_{1-y}$Zr$_y$O$_3$ ceramics diagram: classical ferroelectric BaTiO$_3$-like for $0.00\leq y\leq0.10$,
only one ferroelectric-paralectric transition observed in the range $0.10\leq y\leq0.27$ and relaxor ferroelectric behavior for $0.27< y\leq0.42$.
Then, in 2004, A. Simon {\emph et al.} also
investigated the crossover from a ferroelectric to a relaxor state in lead-free solid solutions in Ref.\cite{BTZ_exp3_simon}, confirming that beyond a definite concentration $y$,
BTZ ceramics show relaxor properties. For BaTi$_{0.80}$Zr$_{0.20}$O$_3$ only one resonance of the permittivity at the ferroelectric-paraelectric $T_C$ of about $315~K$
 is observed, whereas for $y=0.35$ the dielectric anomaly is broad and frequency-dependent as a function of temperature. Moreover in Ref.\cite{BTZ_exp4}, they provide
an EXAFS study of BTZ systems and conclude that BTZ is relaxor and the avarage crystal structure is cubic ($Pm\bar3m$ space group)
in the range $0.25\leq y\leq0.50$. In addition, in a complementary work on BCTZ\cite{BTZ_exp5}, 
the X-ray diffraction has revealed (110) peaks in BaTi$_{0.80}$Zr$_{0.20}$O$_3$ ceramic but a weak tetragonality (i.e. $a/c\sim 1$), that means closeness to the cubic phase. 
However, no experimental data on single crystal samples are available for direct comparison with \emph{ab-initio} results.\\

As for the previous case of (Ba,Ca)TiO$_3$, we first investigate the Ba(Ti,Zr)O$_3$ system first by means of the virtual crystal approximation. 
Then, we go beyond by using supercells-based calculations.

\subsection{Virtual Crystal Approximation (VCA)}
\label{btz_vca}

\subsubsection{Lattice parameters}

We first report the trend of the lattice parameter of the $Pm\bar{3}m$-cubic phase as a function of Zr-composition. Because of the 
bigger volume of BaZrO$_3$ with respect to BaTiO$_3$ (Table~\ref{eigen_bulk}), 
the trend is monotonically increasing with the amount of zirconium and the agreement with the linearity of the Vegard's law is quite 
satisfactory (Fig.~\ref{btz_vca_acell_phonon}(a)).
Contrariwise, because of the bigger ionic radius of Zr than Ti, the tolerance factor\footnote{also in this case, 
we defined an average radius for the $B$-cation as $\overline{r_B}=(1-y)r_{Ti}+(y)r_{Zr}$} decreases without going below 1, that is the value for pure BaZrO$_3$, 
as reported on top of Fig.~\ref{btz_vca_acell_phonon}. 

\subsubsection{Dynamical properties}

\begin{figure}[t]
\centering
\includegraphics[width=\columnwidth]{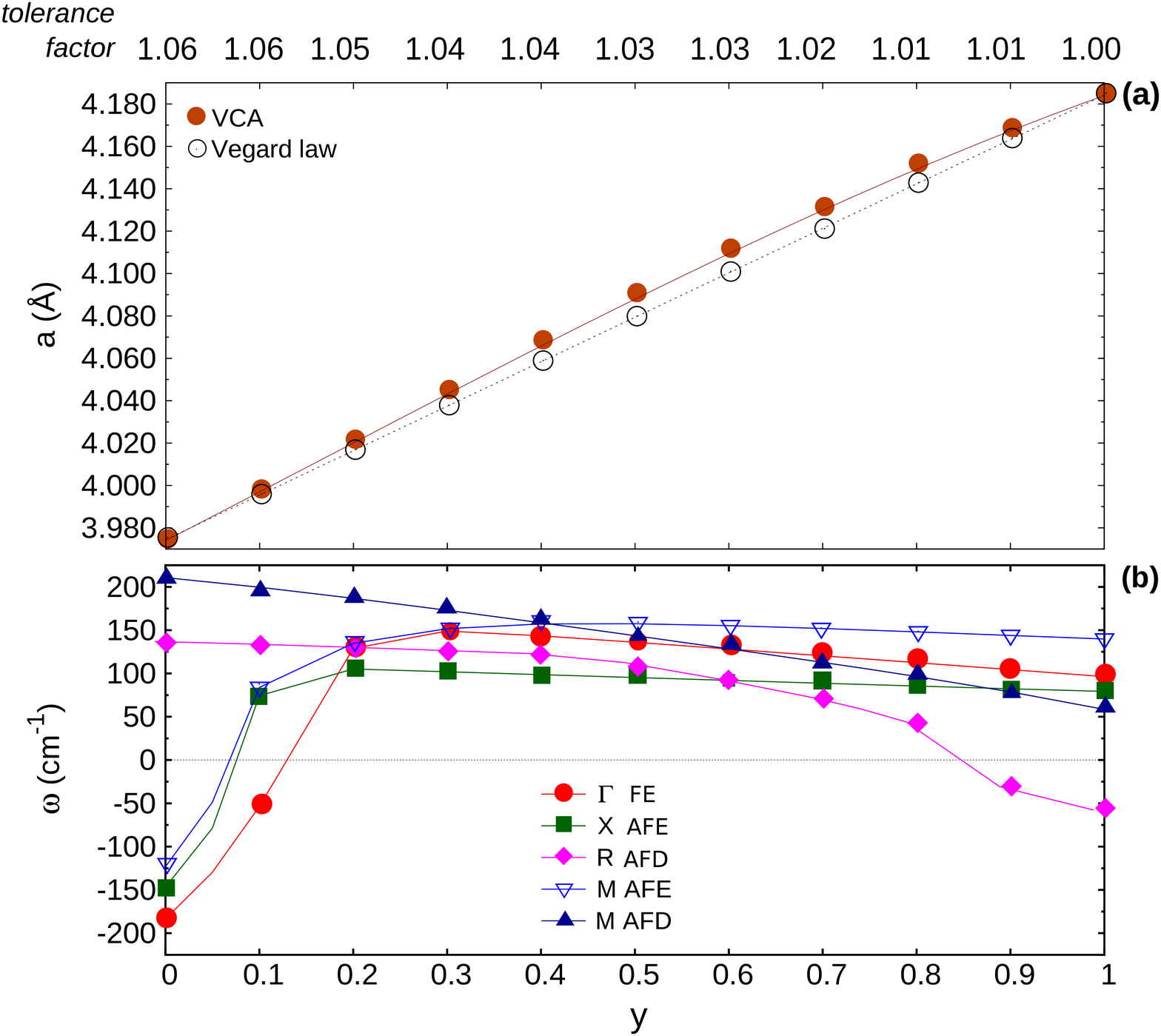}
\caption{\footnotesize (a) Evolution as a function of $y$-composition in BaTi$_{1-y}$Zr$_y$O$_3$ 
of the cubic lattice parameter $a_{cell}$ (in~\AA) relaxed within VCA. Vegard's law has been built on the theoretical values of the 
cubic BaTiO$_3$ and BaZrO$_3$ reported in Table~\ref{eigen_bulk}: $a(y)=(1-y)a_{BTO}+(y)a_{BZO}$. 
(b) Trend of the lowest phonon frequencies (in $cm^{-1}$) at the high-symmetry points of the simple 
cubic Brillouin zone as a function of different $y$-compositions.}
\label{btz_vca_acell_phonon}
\end{figure}

Then, we focus on the dynamics by 
looking at the evolution of the lowest phonon frequencies at the high-symmetry points of the cubic Brillouin zone.
Calculations have been performed on the VCA-optimized $Pm\bar{3}m$ structure. 
It results that the main effect of the Zr-doping is to abruptly stabilize the cubic phase, as clear from Fig.~\ref{btz_vca_acell_phonon}(b). 
In fact, already in the Ti-rich region, i.e. $y>0.10$, the original polar instability vanishes and an AFD unstable mode BaZrO$_3$-like appears at the $R$-point only for $y\geq0.90$.
The reason behind such abrupt weakening of ferroelectricity in the BTZ ``virtual-system" at low $y$-concentration 
can be traced back to abrupt changes in the type of interaction between the virtual cation at the $B$-site and 
the oxygens. In fact, the total IFC along the $B$-O direction is of about $-0.001$~Ha/Bohr$^2$ at $y\geq0.05$. Therefore, the 
 change of sign with respect to the corresponding interaction in pure BaTiO$_3$ (Tables~\ref{ifc_bulk_local}) means that the Zr-substitution strongly favors 
short-range repulsive forces.

Differently from BCT systems, it is not possible to relate such behaviour to the evolution of the tolerance factor or to simple volumetric reasons. 
In fact, in line with the properties of the two parent compounds, $t$ remains larger than one and the volume increases for each intermediate composition. 
The main mechanism involved here is the weakening of the Ti-O interaction when introducing Zr on the same $B$-site, as this corresponds to either 
direct changes in the O $2p$ and $B$ $d$ hydridization mechanism from the electronic perspective or
breaking correlated Ti-O chains necessary to sustain ferroelectricity in BaTiO$_3$ from the lattice dynamics. In fact, we have already seen in Sec.\ref{BZO_phonon}, that the Zr-O interaction in 
BaZrO$_3$ is strongly repulsive. Moreover, differently than the role played by CaTiO$_3$ in BCT, no contribution from the $A$-site is observed, 
as the Ba-O interaction is too weak to sustain alone a polar instability in BTZ.

As a consequence, it is finally not surprising that the ``virtual''-BTZ system does not display any instabilities for a very large range of composition.

\subsubsection{Energetics landscape} 

\begin{figure}[h]
\centering
\includegraphics[width=7cm]{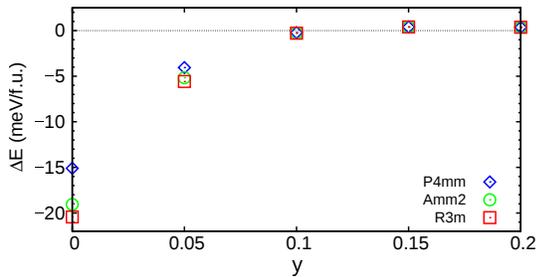}
\caption{\footnotesize Zoom on the variation of the energy gain (in meV/formula unit) of the three polar structures as a function of $y$-composition in the Ti-rich region within VCA.}
\label{btz_vca_energy}
\end{figure}

In line with the drastic changes of the dynamical properties in BTZ predicted by VCA, 
the energy difference both between the three polar phases and with respect to the cubic one is abruptly reduced, as shown in Fig.~\ref{btz_vca_energy}.
In fact, the energy competition is within $0.1~meV/$f.u. at $y=0.10$ and beyond this concentration no stable or 
metastable polar phases are allowed within VCA. 

\subsubsection{Polarization and piezoelectric response}

\begin{figure}[h]
\centering
\includegraphics[width=\columnwidth]{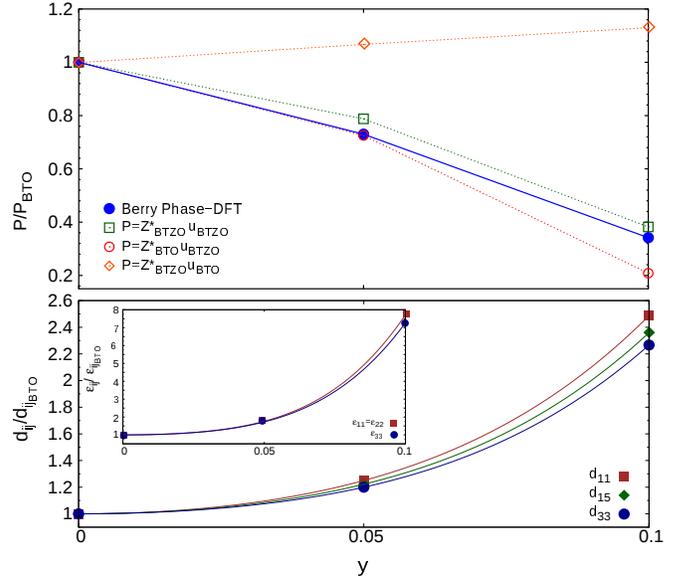}
\caption{\footnotesize Evolution of polarization, free-stress dielectric and piezoelectric coefficients in the rhombohedral-R3m phase 
as a function of $y$-composition in BaTi$_{1-y}$Zr$_y$O$_3$ within VCA.
Values are normalized to the ones of $R3m$-BaTiO$_3$: $P_s^{BTO}\simeq 38~\mu C/cm^{-2}$; $d_{11}^{BTO} \simeq 76$, $d_{15}^{BTO} \simeq 270$ and $d_{33}^{BTO} \simeq 15~ pC/N$; 
$\varepsilon_{11}^{BTO}\simeq 208$ and $\varepsilon_{33}^{BTO}\simeq 11$.
Upper panel: variation of the polarizaton as directly computed by DFT calculations (closed circles) 
and by the phenomenological estimation to disentangle the contribution of varying atomic charges and displacements. Lower panel: trend of the $d_{ij}$
 coefficients and free-stress dielectric response $\varepsilon_{ij}$.}
\label{btz_vca_pol}
\end{figure}

After identifying the abrupt weakening of ferroelectricity in the BTZ ``virtual-system'', we now evaluate the associated effect on the piezoelectric response. 
Therefore, we calculate the spontaneous polarization $P_s$ and the piezoelectric coefficients $d_{ij}$ for the 
stable phase, that is the $R3m$-phase for $y=0.05$ and $0.10$ (Fig.\ref{btz_vca_energy}). Also in this case, as we are mostly interested in the qualitative trend of these 
properties as a function of Zr-doping, we report values normalized to those of $R3m$-BaTiO$_3$. 

In Fig.~\ref{btz_vca_pol} we report values of $P_s$ calculated both via Berry phase (blue closed circles) and 
Born effective charges (green open squares). This is to be compared with the same calculations for BCT in Fig~\ref{bct_vca_pol}. 
In line with the strong weakening of the polar instability, $P_s$ is drastically reduced by the Zr-doping. 
Similarly to BCT, from the futher analysis of the distinguished contribution from effective charge variation and atomic displacements, it results that the 
evolution of $P_s$ has to be widely ascribed to largely reduced atomic motions. However, 
by looking at $y=0.10$, it is important to note that there is a disagreement between the value of polarization 
by fixing the Born effective charges to that of pure $R3m$-BaTiO$_3$ (red open circles) 
or to the ones of $R3m$-BTZ ``virtual-system'' (green open squares) as 
the Zr-doping also affects the Born effective charges. In fact, we have already shown that 
the $Z^*$ associated to Zr is less anomalous than the one of Ti (Table~\ref{eigen_bulk}). 
Therefore, the average effective charge associated to the ``virtual'' $B$-cation is also reduced with increasing Zr-concentration. 
This is an important warning about the ability of VCA in describing the properties arising from the mixing of Zr and Ti on the $B$-site, as they have quite 
different electronic properties (see Sec.\ref{vcaVSsc_btz} below).

Such $P_s$, actually, favors piezoelectricity. In fact, the huge increase of dielectric response, $\varepsilon_{ij}$, 
associated to the drastic softening of the polar mode,
largely compensates the decrease of $P_s$. This results in the enhancement of the piezoelectric response with respect to pure BaTiO$_3$,  
as reported in Fig.~\ref{btz_vca_pol}. Accordingly, the values of the lowest-frequency polar mode in the stable $R3m$-phase evolve like 169, 146, 69 (in $cm^{-1}$) 
for $y=0.00, 0.05$ and $0.10$, respectively.\\

Despite the overall predictions provided by VCA seem to be reliable when comparing the behaviour of the two parent compounds and 
with respect to experiments, supercells-based calculations are better required to clarify the local effect of Zr-doping on the dynamics of BaTiO$_3$.

\subsection{Supercell Calculations} 
\label{btz_super}

The dynamical properties of BaTi$_{0.875}$Zr$_{0.125}$O$_3$ and BaTi$_{0.50}$Zr$_{0.50}$O$_3$ supercells are here investigated. Structural details are provided
 in Sec.~\ref{supercell_description}

\begin{figure}[b]
\centering
\includegraphics[width=8cm]{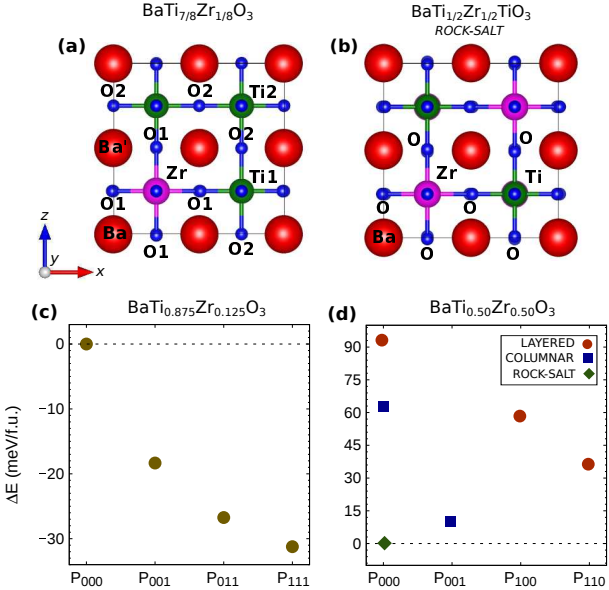}
\caption{\footnotesize Schematic representation of the atomic arrangement in BaTi$_{0.875}$Zr$_{0.125}$O$_3$ (a) and in rocksalt-BaTi$_{0.50}$Zr$_{0.50}$O$_3$(b).
 Labels help the visualization of the main atomic pair analyzed in Tables~\ref{ifc_BTZ_super_12} and~\ref{ifc_BTZ_50_cubic}.
Energy lowering (in meV/f.u.) with respect to the highy-symmetry $Pm\bar3m$ phase for BaTi$_{0.875}$Zr$_{0.125}$O$_3$ (c) and with respect to highy-symmetry rocksalt-$Fm\bar3m$
configuration for BaTi$_{0.50}$Zr$_{0.50}$O$_3$ (d) as a function of different optimized polar states by means of supercells.
The $x$-axis notation refers to the direction of the components of polarization.}
\label{structure_BTZ}
\end{figure}

\subsubsection{BaTi$_{0.875}$Zr$_{0.125}$O$_3$}

\begin{table*}[t]
\footnotesize
\centering
\begin{tabular}{|c||c|c|c|c||cc|c|c|c|}
\hline
\multicolumn{10}{|c|}{\bf BaTi$_{0.875}$Zr$_{0.125}$O$_3$} \\
\hline
 & \multicolumn{4}{c||} {\bf VCA} & \multicolumn{5}{c|} {\bf 2x2x2 CELL} \\
\hline
Type & Direction &  Total &   DD    &   SR &  Atoms &                         &   Total &   DD    &   SR  \\
\hline
$A-A'$ & ($\parallel$) & -0.011 & -0.006 & -0.005 & Ba-Ba'  & ($\parallel$) &  -0.012 & -0.006 & -0.0061 \\
       & ($\perp$)     & +0.004 & +0.003 & +0.001 & d=4.043 \AA   & ($\perp$) &  +0.004 & +0.003 & +0.001 \\
       & d=3.997\AA &  &  &            &                           &  &   &  &  \\
\hline
$B-B'$ & & & &                                        &Ti$_1$-Ti$_2$  & ($xx'$) & -0.068 & -0.038 & -0.029 \\
       & ($\parallel$) & -0.067 & -0.037 & -0.030    & d=4.002 \AA  & ($yy'$) & +0.008 & +0.018 & -0.010 \\
       & ($\perp$) &     +0.009 & +0.018 & -0.011    &               & ($zz'$) & +0.008 & +0.019 & -0.011 \\[4pt]
       & d=3.997 \AA & & &                         & Ti$_1$-Zr       & ($xx'$) & -0.069 & -0.034 & -0.034  \\
       & & &                                      &  &  d=4.002 \AA   & ($yy'$) & +0.008 & +0.019 & -0.011 \\
       & & & &                                    &  & ($zz'$) & +0.008 & +0.019 & -0.011 \\
\hline
$A-B$ & & & &                                         & Ba-Ti$_1$  & ($xx'$) & -0.028 & -0.022 & -0.007 \\
      & ($\parallel$) & -0.030 & -0.023 & -0.008     & d=3.478 \AA   & ($yy'$) & +0.014 & +0.010 & +0.003 \\
      & ($\perp$)     & +0.014 & +0.011 & +0.003     &                & ($zz'$) & +0.013 & +0.011 & +0.002 \\[4pt]
      & d=3.461  \AA & & &                           & Ba-Zr      & ($xx'$) & -0.0312 & -0.0200 & -0.0113 \\
      & & &                                        &   & d=3.502 \AA   & ($yy'$) & +0.013 & +0.011 & +0.002 \\
      & & & &                    &                    & ($zz'$) & +0.013 & +0.011 & +0.002 \\
\hline
$B-O_1$  & & & &                                         & Ti$_1$-O$_1$  & ($\parallel$) & -0.032 & +0.239 & -0.271 \\
         & & & &                                         & d=1.929\AA   & ($\perp$)     & -0.015 & -0.052 & +0.037 \\[4pt]
         &  &  &  &                                     & Zr-O$_1$      & ($\parallel$) & -0.039 & +0.199 & -0.238 \\
         & ($\parallel$)     & -0.009 & +0.229 & -0.238    & d=2.073\AA   & ($\perp$)     & -0.013 & -0.040 & +0.027 \\[4pt]
         & ($\perp$)     & -0.020 & -0.045 & +0.025    & Ti$_1$-O$_2$  & ($xx'$) & +0.013 & +0.239 & -0.226 \\
         & d=1.998  \AA & & &                                         & d=2.012 \AA   & ($yy'$) & -0.023 & -0.039 & +0.016 \\
         & & & &  &                                                       & ($zz'$) & -0.021 & -0.046 & +0.025 \\[4pt]
         & & & &                                         & Ti$_2$-O$_2$  & ($\parallel$) & +0.003 & +0.244 & -0.241 \\
         & & & &                                        & d=1.990 \AA   & ($\perp$)     & -0.021 & -0.043 & +0.022 \\
\hline
$A-O_1$ & ($xx'$) & -0.001 & +0.013 & -0.014   & Ba-O$_1$  & ($xx'$) & +0.000 & +0.012 & -0.0113 \\
        & ($yy'$) & -0.005 & -0.006 & +0.002   & d=2.860 \AA  & ($yy'$) & -0.010 & -0.015 & +0.005 \\
        & ($zz'$) & -0.011 & -0.016 & +0.005   &               & ($zz'$) & -0.004 & -0.006 & +0.002 \\
        & d=2.826  \AA & & &  &                  & & & &  \\

\hline
\end{tabular}
\caption{\footnotesize Interatomic force constants from VCA and supercell-based calculations for BaTi$_{0.875}$Zr$_{0.125}$O$_3$. 
Units are in Ha/Bohr$^2$. Directions $xx'~(\parallel)$, $yy'~(\perp)$ and  $zz'~(\perp)
$ refer to local coordinates system of the different pairs of atoms. 
Distances (in \AA~) between the selected atoms are also reported. Atoms' notation in the second column is consistent with Fig.~\ref{structure_BTZ}(b). }
\label{ifc_BTZ_super_12}
\end{table*}

BaTi$_{0.875}$Zr$_{0.125}$O$_3$ supercell (Fig.~\ref{super_12}(e))
hosts several instabilities in the  cubic $Pm\bar{3}m$ phase, all related to ferroelectric and antiferroelectric distortions.
The most unstable, with $\omega\simeq 214i~cm^{-1}$, is a polar mode mostly related to the opposite motion of Ti and O, 
while the Zr atom slightly moves in the same direction as the oxygens. 
The difference in these relative displacements affects the neighboring atom lying on the polar direction, that 
slightly moves with the oxygens as well. 
That is, in terms of the labels indicated in Fig.~\ref{structure_BTZ}(a), if the polar distorion is along the $x$ direction, for instance, Ti$_1$ slightly off-centers with Zr 
in the same direction as the oxygens, but opposite to the other Ti-atoms. Ba-atoms are almost fixed as in the parent BaTiO$_3$. 
Such different dynamics is made clear when looking at the interatomic force constants reported in Table~\ref{ifc_BTZ_super_12}. 
In fact, the original destabilizing Ti-O interaction of BaTiO$_3$ is not preserved for each direction in the system. 
The Ti-atoms having Zr as first-neighbors experience a strong repulsive Ti-O interaction along the O-Zr-O-Ti-O chain direction, 
while the long-range forces remain the major ones 
along the other directions as well as for the other Ti-atoms. 
This is strongly ascribed to the ``chain-like" character of the polar instability of BaTiO$_3$, as discussed in Sec.~\ref{BTO_phonon}:
 the persistence of the polar instability requires a minimum number of correlated Ti-O displacements \cite{phil2} that, however, are prevented along the chain containing Zr. 
In fact, according to the properties of BaZrO$_3$ described in Sec.~\ref{BZO_phonon}, the Zr-O interaction remains strongly repulsive also in this Ti-rich solid solution. 
However, along the preserved O-Ti-O-Ti-O chains, the long-range interaction is even indirectly strenghted by the Zr-presence. In fact, the bigger size of Zr 
(i.e. of the [ZrO$_6$]-octahedral volume) makes the Ti-O bond's length perpendicular to the O-Zr-O-Ti-O chain direction longer than in pure BaTiO$_3$. This mechanism 
locally favors the long-range interaction. 
Accordingly, the ``on-site" force constants tensor associated to the Ti-atoms adjacent to Zr are anisotropic: by referring to Ti1 of Fig.~\ref{structure_BTZ}(a), 
the ``on-site" terms are $(+0.247, +0.132, +0.132)$~Ha/Bohr$^2$. Therefore, they are much more stiffer along the Zr-direction, while softer along the tranversal directions than in 
BaTiO$_3$ (see Table~\ref{sfc_bulk}). The other Ti-atoms display a slightly anisotropic tensor, but similar to the one in BaTiO$_3$: by referring to Ti2 of Fig.~\ref{structure_BTZ}(a), 
values are $(+0.162, +0.151, +0.162)$~Ha/Bohr$^2$.
In contrast, Zr displays an isotropic tensor BaZrO$_3$-like with diagonal values of about $+0.229$~Ha/Bohr$^2$,
as it experiences the same interactions in each direction.
Concerning Ba at the $A$-site, the dynamics is the same as in pure BaTiO$_3$. In fact, the $A$-O interatomic force constant is close to zero along parallel direction of the coupling meaning 
that Ba distortion does not lead ferroelectricity in BaTi$_{0.875}$Zr$_{0.125}$O$_3$. The $A$-$B$ interaction is also similar to the one in the two parent compounds 
reported in Table~\ref{ifc_bulk_local}.

Such complex dynamics results in slightly reduced total polarization and more spread energy landscape with respect to the case of the parent BaTiO$_3$.
On one hand, the calculated spontaneous polarization is of about $28~\mu C/cm^2$ and $35~\mu C/cm^2$ for the $P4mm$ and $R3m$ phases, respectively, 
while it is of about $34~\mu C/cm^2$ and $38~\mu C/cm^2$ for the corresponding phases in pure BaTiO$_3$. This has to be assigned to the 
local depolarizing contribution arising from the O-Zr-O-Ti-O chain.
On the other hand, the energy gain of the three polar phases relative to the cubic phase is higher than in BaTiO$_3$ and it appears much more pronounced for the 
$Amm2$ and $R3m$ phases as reported in Fig.~\ref{structure_BTZ}(c). This effect traces back to the overall volume increase, 
that further favors the rhombohedral phase. 

\subsubsection{BaTi$_{0.50}$Zr$_{0.50}$O$_3$}
\label{btz50_solution}

\begin{table*}[t]
\footnotesize 
\centering
\begin{tabular}{|c|c|c|c|c|c|c|}
\hline
\multicolumn{7}{|c|}{\bf BaTi$_{0.5}$Zr$_{0.5}$O$_3$ ($\bm {P4mmm$}-SUPERCELL)} \\
\hline
&  & & &  & & \\
Configuration & $\Gamma (0~0~0) $  &  Z (0~0~$\frac 12$)  & M ($\frac 12$~$\frac 12$~0)& A ($\frac 12$~$\frac 12$~$\frac 12$) & R (0~$\frac 12$~$\frac 12$~) & X (0~$\frac 12$~0)\\
&  & & & & &  \\
\hline
{\bf layered}  & 274.18i & 271.42i  & 65.17i  &  60.85i & 259.90i  & 262.65i \\
{\bf [110]}  &     &  &  &  &   &  \\
\hline
{\bf columnar}  & 304.05i & 90.60 & 301.88i  & 75.43  & 76.32  & 299.4i  \\
{\bf [001]}  &    &    &      &    &    & \\
\hline
\end{tabular}
\caption{\footnotesize Lowest phonon frequencies ($cm^{-1}$) at the high symmetry points of the tetragonal $P4mmm$ Brillouin zone.
As BaTi$_{0.5}$Zr$_{0.5}$O$_3$ has no instabilities within VCA, only standard DFT results are reported.}
\label{BTZ_50per_phonon}
\end{table*}

\begin{table}[t]
\centering
\footnotesize
\begin{tabular}{|c|cc|c|}
\hline
\multicolumn{4}{|c|} {\bf 2x2x2 CELL - \bm {$Fm\bar{3}m$}} \\
\hline
\multicolumn{4}{|c|}{\bf BaTi$_{0.50}$Zr$_{0.50}$O$_3$} \\
\hline
Type & Atoms &                         &   Total   \\
\hline
$B-B'$ & Ti-Zr          & ($\parallel$) & -0.063  \\
       & d=4.075 \AA   & ($\perp$)     & +0.008 \\

\hline
$A-B$ & Ba-Ti          & ($\parallel$) & -0.027 \\
      & d=3.529 \AA   & ($\perp$) & +0.013  \\[4pt]
      & Ba-Zr          & ($\parallel$) & -0.030  \\
      & d=3.529 \AA   & ($\perp$) & +0.0130 \\
\hline
$B-O$     & Ti-O          & ($\parallel$) & -0.002  \\
          & d=1.978 \AA  & ($\perp$)     & -0.020 \\[4pt]
          & Zr-O          & ($\parallel$) & -0.025 \\
          & d=2.097 \AA  & ($\perp$)     & -0.015 \\
\hline
$A-O$     & Ba-O          & ($xx'$) & +0.002  \\
          & d=2.882 \AA  & ($yy'$)  & -0.011 \\
          &              & ($zz'$)  & -0.005 \\
\hline
\end{tabular}
\caption{\footnotesize Interatomic force constants from supercell calculations for BaTi$_{0.50}$Zr$_{0.50}$O$_3$ in the cubic-$Fm\bar{3}m$ phase.
Units are in (Ha/Bohr$^2$). Directions $(\parallel), (xx')$ and $(\perp), (yy'),(zz')$ refer to the local coordinates system between different atomic coupling.
Distances in \AA~between the selected atoms are also reported. Atomic labels refer to Fig.~\ref{structure_BTZ}(b).}
\label{ifc_BTZ_50_cubic}
\end{table}

The case of BaTi$_{0.50}$Zr$_{0.50}$O$_3$ (Fig.~\ref{super_12}(f-h)) clearly reveals the importance of the atomic ordering in the dynamical properties. 
First result is that the arrangement with the lowest energy is the high-symmetry-$Fm\bar{3}m$ rocksalt 
configuration compared to the ordered-$P4mmm$ structures based on chains or layers of same type of $B$-cations. 
The three structures have been optimized and the energy lowering with respect to the highest energy layered configuration
 is of about $-30.95~meV/$f.u. for the columnar configuration and $-93.73~meV/$f.u. for the rocksalt. 

Beyond that, all the three configurations show very different dynamical properties. In fact, by looking at the phonon frequencies reported 
in Table~\ref{BTZ_50per_phonon}, unstable modes appear at each high symmetry point of the $P4mmm$-tetragonal Brillouin zone of the layered-based supercell. 
Specifically, the two instabilities appearing at the $M$ and $A$ points are associated to antiferrodistortive modes due to oxygen rotations, while all the 
other instabilities are related to the polar instability arising from $\Gamma$ and mostly ascribed to the (Ti,O)-polar motion along the direction of preserved O-Ti-O-Ti-O. 
Zr displays the same dynamics as described for the BaTi$_{0.875}$Zr$_{0.125}$O$_3$ case.
Within the columnar configuration, the 
instabilities appear only along the $\Gamma-X-M$ line and are due to the polar distortion arising from the $\Gamma$-point with the same character as the former cases. 
In sharp contrast, no instabilities appear within the rocksalt configuration. In particular, the lack of correlated O-Ti-O-Ti-O in any direction makes the cubic phase stable. 
As reported in Table~\ref{ifc_BTZ_50_cubic}, in fact, both the Ti-O and Zr-O interactions are dominated by the short-range forces and 
the slightly destabilizing interaction between Ba and O atoms is not enough to globally destabilize the system. 

Since the Zr-doping directly affects the $B$-O interaction, the system clearly prefers to keep an isotropic surrounding environment      
in order to preserve the same kind of interaction in each direction, i.e.
O-Zr-O-Ti-O chains like in the $Fm\bar{3}m$-structure. This arrangement prevents the original Ti-driven polar distortion of BaTiO$_3$.
Therefore, the ground-state for the $y=0.50$ composition is globally non-polar, as reported in Fig.~\ref{structure_BTZ}(d).

\subsection{VCA vs SUPERCELL calculations}
\label{vcaVSsc_btz}

By comparing the results reported in Secs.~\ref{btz_vca} and~\ref{btz_super}, it becomes quite clear that the VCA method is not suitable for the Ba(Ti,Zr)O$_3$ solid solutions. 
The decrease in the polarization with increasing the Zr-concentration is obtained in both cases. However, within VCA, neither the energetics nor the dynamical properties are well reproduced. 
In particular, VCA does not detect any instability in the cubic phase already for $y>0.10$, whereas it is not the case as found out in BaTi$_{0.875}$Zr$_{0.125}$O$_3$-supercell 
as well as in the experiments introduced at the beggining of the present Section. 
As already widely discussed, the reason behind such failure is that VCA does not allow to access distinct contributions to the dynamics arising from the different nature of the 
mixed cations as well 
as the effects due different atomic ordering. The latter is, in fact, the main variable controlling the dynamics in BTZ. 
Additionally, we know that the FE instability in BaTiO$_3$ is strongly sensitive to the O $2p$ - Ti $3d$ hybridization. 
Therefore, as Zr occupies $4d$ states, the changes in the electronic properties induced by the Zr-doping directly affect the dynamics of the system. 
As such, the way VCA combines the different electronic properties from the parents BaTiO$_3$ and BaZrO$_3$ is also determinant in the failure of the VCA approach for BTZ.
 
\begin{figure}[t]
\centering
\includegraphics[width=\columnwidth]{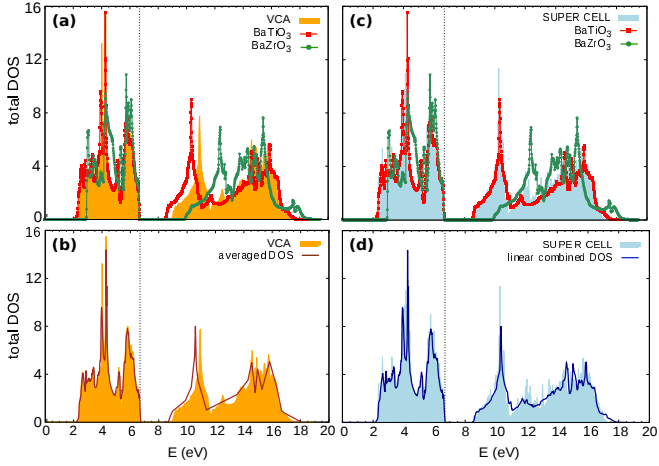}
\caption{\footnotesize Calculated density of states for BaTiO$_3$ (red), BaZrO$_3$ (green) and BaTi$_{0.875}$Zr$_{0.125}$O$_3$ (orange). Upper panels:
 DOS of the BTZ solid solution as obtained within VCA (left) and supercell calculation (right) in comparison with the parent compounds.
Bottom panels: empirical reproducion of DOS as obtained
from horizontal average to reproduce the VCA (left) and linear combination for the supercell case (right). }
\label{dos}
\end{figure}

In Fig.~\ref{dos} we compare the electronic density of states (DOS) within the two approaches. 
It appears that VCA acts in making an ``horizontal'' average of the DOS of the parent compounds. 
By aligning the upper part of the valence bands to that of BaTiO$_3$, this is specially evident for the conduction bands as the energy band gap of BaZrO$_3$ is larger than BaTiO$_3$, 
as reported in Fig.~\ref{dos}(a,c).
In fact the DOS calculated by VCA is 
 shifted towards the bottom of BaZrO$_3$ conduction proportionally to the 12.5\% composition of zirconium 
as it is linearly interpolated between the $3d$ states of Ti and $4d$ states of Zr. 
In panel~\ref{dos}(b), we reproduce this average by adding up the density of peaks facing one with the other with the weight fixed by the chosen composition.
The result, plotted on top of the VCA graph, reproduces nicely the calculated DOS (non-linear effects play a minor role in the creation of fictitious virtual atom).

The way supercell-based calculations combine properties from the parent compounds is significantly different.
It results to be a straighforward linear combination of the two densities. In our example, we just added up $\frac78 DOS_{BTO}+\frac18 DOS_{BZO}$ and the plot
extraordinarily overlaps the calculated one.
This distinct way of averaging the electronic properties can be the reason why the two approaches reproduce different ground-state for the same
composition. The incorrect average location of the $d$-states of the virtual atoms in VCA (i.e. at higher energy) prevents the correct $p$-$d$ hybridization, 
which is at the basis of the ferroelectric distortion in BaTiO$_3$.  

\subsection{Role of cations arrangement: BaZrO$_3$/BaTiO$_3$ supercells}
\label{superlattice}

\begin{figure}[b]
\centering
\includegraphics[width=8cm]{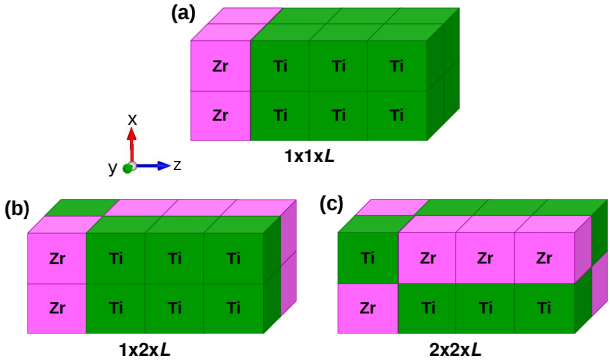}
\caption{\footnotesize Skeleton of the investigated BZO/$m$BTO-based supercells. (a) 1x1x$\mathit{L}$ superlattice based on one single chain of BaZrO$_3$-$m$BaTiO$_3$
with decreasing composition of zirconium. The smallest chain 
with $L$=2 corresponds to the layered 50\%-composition. The represented chain is doubled along the $x$ and $y$ direction to help the visualization of the 
crystal periodicity.  (b)-(c) 1x2x$\mathit{L}$ and 2x2x$\mathit{L}$ supercells 
based on alternating chains BaZrO$_3$-$m$BaTiO$_3$ and BaTiO$_3$-$n$BaZrO$_3$ to keep a 50\% global composition. 
The smallest cells with $L$=2 correspond 
to the columnar and Rock-Salt 50\%-composition, respectively. The represented 1x2x$\mathit{L}$-case is doubled along the $x$ to help visualization. }
\label{super_chain}
\end{figure}

Since the dynamics in Ba(Ti,Zr)O$_3$ is strongly related to the geometrical and atomic ordering while the polar distortion in BaTiO$_3$ requires correlated 
displacements along (Ti,O)-chains, 
we find it interesting to explore more deeply different compositions and cations arrangements.
Thus, we have performed additional calculations for idealized BaZrO$_3$/BaTiO$_3$ (BZO/BTO) supercells (Fig.~\ref{super_chain}) in order to clarify 
the role played by the geometrical environment and composition in getting ferroelectricity and homogenous polarization in 
Ba(Ti,Zr)O$_3$ solid solutions.

At first, we considered a given structural ordering with different composition.
We built 1x1x$\mathit{L}$ superlattices (BZO/$m$BTO) by adding
up to six layers of BaTiO$_3$ to one unit cell thick layer of BaZrO$_3$, as represented in Fig.~\ref{super_chain}(a).  
This corresponds to decreasing concentrantion of zirconium from 50$\%$ for $L$=2, to 33$\%$ for $L$=3, 25$\%$ for $L$=4, 20$\%$
 for $L$=5 and 17$\%$ for $L$=6.
The total lenght in terms of unit cells is $L=m+1$.
Then, proceeding from the evolution with $\mathit{L}$ of the energetics and polarization along the stacking direction,
we tried to explain the appearence of ferroelectricity by a simple electrostatic model.

\begin{figure}[t]
\centering
\includegraphics[width=\columnwidth]{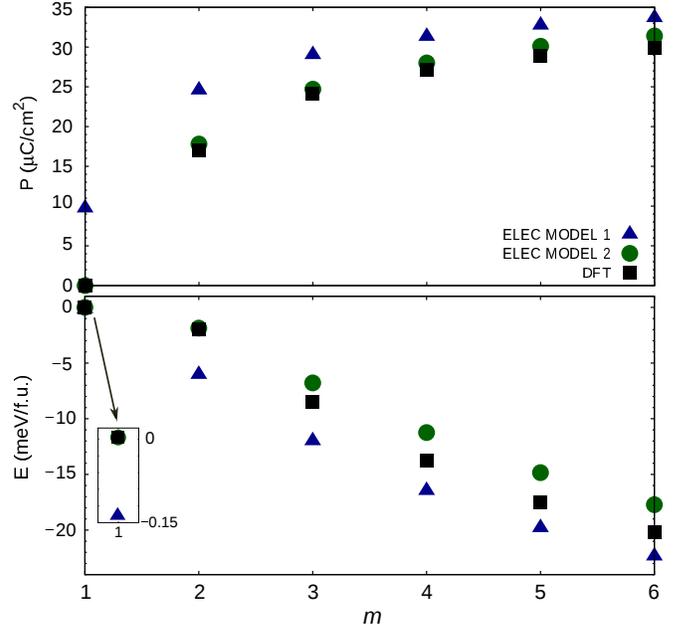}
\caption{\footnotesize Evolution of the total polarization (in $\mu C/cm^{2}$) and energy (in meV/formula unit) as a function of increasing number $m$ of BTO-layers.
Values obtained from DFT calculations and from the electrostatic model are reported. We refer to results coming from the 
first approach based on the minimization at fixed $P$ as model 1, while to those based on the relaxed patter as model 2. 
The inset shows the different energies for $m=1$. The $E$ from DFT calculations refers to the relative energy gain between the
optimized ferroelectric phase and the centro-symmetric reference with volume fixed to that of the polar structure.}
\label{pol_superlatt}
\end{figure}

In detail, we started our study by investigating the appearance of polarization along the epitaxy direction of the 
1x1x$\mathit{L}$ superlattice. 
For each lenght $L$, we fully relaxed the centro-symmetric 
$P4/mmm$ and the polar $P4mm$ structures, where the inversion symmetry breaking is due to the atomic displacements along 
the $z$ direction. As expected, we found no polar distortions along the $z$ direction for $L$=2. Because of the periodicity of the crystal, this configuration 
corresponds, in fact, to the layered structure described in Sec.~\ref{btz50_solution}, Fig.~\ref{super_12}.
Conversely, for $L\geq$3 the polar structures have an energy gain with respect to the paraelectric one and values 
of polarization increase with the number $m$ of BaTiO$_3$ layers. Values of $P_s$ are reported in Table~\ref{PdvsL} 
and in Fig.~\ref{pol_superlatt}. 
Then we described the atomic distortion from the paraelectric to 
the ferroelectric phase as a cooperative motion around the center of mass (CM) of the system, in order to better clarify 
the driving mechanism. This description has revealed the active role of Zr-atom
 in the polarization against the dynamical properties found out in bulk BaZrO$_3$ as well as in the investigated 
Ba(Ti,Zr)O$_3$ solid solutions (Sec.~\ref{BZO_phonon} and \ref{btz_super} respectively). 
For $m\geq2$, BZO/$m$BTO superlattice behaves like 
BaTiO$_3$ with the opposite motion of the O-anions and (Ti,Zr)-cations. 
In terms of stiffness, the zirconium atom experiences a much smaller ``on-site'' force constant in the direction of the epitaxy, 
{\it i.e.} of the Ti-atoms, 
that allows for the Zr-polar motion.
Moreover, the associated Born effective charge, of about $\simeq 7.2~e$, is considerably anomalous along that direction and comparable to that of Ti.

For sake of completeness, we also performed DFPT calculations in order to evaluate the $d_{33}$ piezoelectric coefficient in the $P4mm$ phase. 
Results are reported in Table~\ref{PdvsL}. 
It is noteworthy 
that this polar phase is not the ground-state. In fact, it still presents 
phonon polar instabilities (not shown) in the ($x,y$) plane, i.e. along the preserved Ti-O-Ti-O chains,
with dynamical properties similar to the Ba(Ti,Zr)O$_3$ solid solutions described in Sec.~\ref{btz_super}. \\

\begin{table}[t]
\centering
\footnotesize
\begin{tabular}{|c|c|c|c||c|c|c||c|c|c|}
\hline
  & \multicolumn{3}{c||}{{\bf1x1x}$\mathbf{L}$} & \multicolumn{3}{c||}{{\bf1x2x}$\mathbf{L}$} & \multicolumn{3}{c|}{{\bf2x2x}$\mathbf{L}$} \\
\hline
$\mathbf{L}$ & $\Delta$E & P & d$_{33}$ & $\Delta$E & P & d$_{33}$ &  $\Delta$E & P & d$_{33}$ \\
\hline
2 & 0.0  & 0.0  & 0.0  & 0.0   & 0.0  & 0.0  & 0.0   &  0.0 & 0.0 \\
3 & -1.0 & 17.0 & 77.3 & -0.8  & 10.2 & 34.0 & -1.3  &  11.1 & 28.0 \\
4 & -4.2 & 24.2 & 57.2 & -5.6  & 15.9 & --   & -9.2  &  17.3 & -- \\
5 & -6.7 & 27.2 & 52.3 & -10.2 & 18.1 & --   & -16.3 & 19.7 & -- \\
6 & -8.6 & 28.9 & 49.7 & -13.8 & 19.3 & --   & -21.7 & 21.0 &  -- \\
\hline
\end{tabular}
\caption{\footnotesize Comparison of P (in $\mu C/cm^2$) 
and d$_{33}$ (in $pC/N$)
as a function of the lenght $L$ in the three investigated atomic arrangements, as obtained by Berry-phase and DFPT calculations. 
For 1x2x$\mathit{L}$ and 2x2x$\mathit{L}$ 
supercells, DFPT calculations have been performed only for $L$=3. 
For comparison, in $P4mm$-BaTiO$_3$ we obtained $d_{33}\simeq 43.3~pC/N$.
Relative energy gain between the optimized ferroelectric 
and paraelectric phases ($\Delta$E(FE-PE) in meV/f.u.) is also reported.}
\label{PdvsL}
\end{table}

In order to interpret the latter results, specifically the polar activation of Zr, here we adapt the simple model proposed in Ref.[\cite{phil_handbook}]
describing the behavior of dielectric/ferroelectric multilayers to our BZO/$m$BTO superlattice.

Neglecting interface corrections, the total energy of a ($1,m$) superlattice can be written as
\begin{align}
E&(P_{\footnotesize BZO},P_{BTO};m) = U_{BZO}(P_{BZO})+mU_{BTO}(P_{BTO}) \nonumber\\ 
&+C(m)(P_{BTO}-P_{BZO})^2
\label{E_tot_complete}
\end{align}
Here $P$ is the polarization arising from the displacement of the ions from their high-symmetry positions 
under the condition of zero electric field,
$U$ is the internal energy of bulk BaZrO$_3$ and BaTiO$_3$ at zero field as a function of $P$ and $C(m)(P_{BTO}-P_{BZO})^2$ is
macroscopic electrostatic energy, $E_{elec}$, resulting from the eventual presence of non-vanishing electric fields in the layer when $P_{BZO}$ and $P_{BTO}$ are different.
This term typically acts as an energy penalty which tends to reduce the polarization mismatch in polarizing the dielectric layer and depolarizing the ferroelectric one. 
In practice, when the dielectric layer is sufficiently polarizable, this term forces the system to adopt a uniform polarization along the stacking direction 
($z$ direction in Fig.~\ref{super_chain}), i.e. $P_{BZO}=P_{BTO}=P$. 
In this case the model reduces to \cite{rabe_superlattice,rabe_sto_pto_superlattice,eric_superlattice}
\begin{equation}
E(P; m) = U_{BZO}(P)+mU_{BTO}(P)
\label{E_tot}
\end{equation} 

The energies $U_{BTO}(P)$ and $U_{BZO}(P)$ can be directly obtained from appropriate DFT calculations on bulk compounds.
In the case of BaTiO$_3$, we built the adiabatic
 path from the paraelectric to the ferroelectric $P4mm$ phase (discussed in Sec.~\ref{bulk_energy}) by 
means of linear interpolation of the atomic displacements. During this interpolation the volume is fixed to the one of the polar structure. For each intermediate configuration,
we computed the internal energy and polarization. 
This results into a double-well energy profile that we fit with the standard polynomial expansion
\begin{equation*}
U_{BTO}(P) \simeq \alpha_T P^2+\beta_T P^4+\gamma_T P^6
\end{equation*}
where $\alpha_T$, $\beta_T$, $\gamma_T$ are the fitting parameters. 
\footnote{The sixth order term is introduced to ensure a proper description of the physics of the system as 
it is close to a tricritical point, i.e. change of the order of the phase transition 
via a change in sign of the coefficient of the fourth order term.}
We used a similar expansion for BaZrO$_3$. However, according to the paraelectric nature of this compound, 
the coefficient of the second order term is positive. 
Thus we restricted the expansion to the fourth order and we write
\begin{equation*}
U_{BZO}(P) \simeq \alpha_Z P^2+\beta_ZP^4.
\end{equation*}

\begin{figure}[t]
\centering
\includegraphics[width=7.5cm]{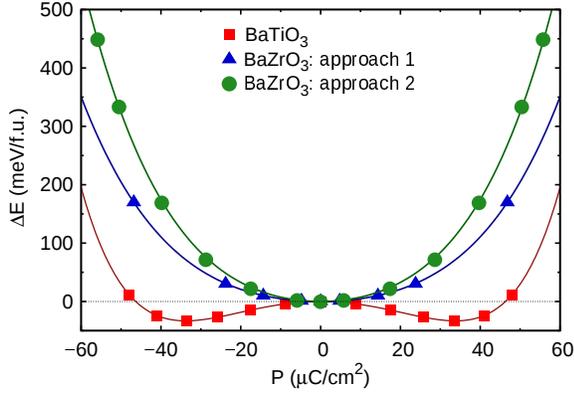}
\caption{\footnotesize Variation of total energy (in meV/f.u.) as a function of the varying polarization (in $\mu C/cm^{2}$) from the centric to the polar configurations 
for distorted bulk BaTiO$_3$ (red squares) and BaZrO$_3$. For the latter, curves as obtained within the two approaches are reported: 
(blue tringles) from the minimization procedure at fixed $P$; (green circles) from the relaxed pattern of distortion of 1x1x3 superlattice. }
\label{E_BTO_BZO}
\end{figure}

In this case, as the system does not show any polar instability, 
 the only way to follow the evolution of the internal energy $U_{BZO}(P)$
with the ferroelectric distortion is to freeze
the pattern of the motion along the $z$ direction.
We defined the total distortion $\tau$ as a linear combination of the eigendisplacements associated to the polar modes,
\emph{i.e.} $\tau=a_1\eta_{TO1}+a_2\eta_{TO2}+a_3\eta_{TO3}$, with $\sum_{i}a_i^2=1$ and $\langle\eta_{i}|M|\eta_{i}\rangle=1$.
The first approach was to determine the coefficients of the expansion by minimizing $U_{BZO}(P)$
to equilibrium values of $\eta_{i}$ with the constrain of fixed $P$, as proposed in Ref.~\cite{sai_rabe}.
From this minimization we obtained $a_1$=0.945, $a_2$=0.325 and $a_3$=0.044, that corresponds to a pattern dominated
by the softest $TO_1$ mode of BaZrO$_3$ driven by the Ba motion (see Table~\ref{eigen_bulk}).
However, this result is not in full agreement with the pattern of distortion obtained from the atomic relaxation in the superlattice
 previously described. 
Therefore, we also built $U_{BZO}(P)$ from the relaxed pattern. In detail, we considered the distorsion of the BaZrO$_3$
unit cell in the 1x1x3 superlattice and we projected it on each $TO_i$ mode in order to get the $a_i$ coefficient to be
compared with the previous procedure. We obtained $a_1$=0.595, $a_2$=0.802 and $a_3$=0.051, that clearly reveal
the key role of zirconium.
The $U_{BZO}(P)$ curves obtained via the two procedures are shown in Fig.~\ref{E_BTO_BZO}.

Then, from the minimization of Eq.~\ref{E_tot} with respect to $P$, we obtain that a spontaneous polarization is admitted 
if the condition $m>\frac{\alpha_Z}{|\alpha_T|}$ is satisfied, otherwise $P$ is zero.
The resulting formula for the polarization and energy are
\begin{eqnarray*}
P^{2}(\alpha,\beta,m) & \simeq & -\frac{m\alpha_T+\alpha_Z}{2(m\beta_T+\beta_Z)}  \\
E(\alpha,\beta, m) & \simeq & -\frac{(m\alpha_T+\alpha_Z)^2}{4(m\beta_T+\beta_Z)} 
\end{eqnarray*}

Coefficients are reported in Table~\ref{fit_parameters}. 
The evolution of $P$ and $E$ resulting from the electrostatic model within the two different 
approaches and from DFT calcualtions are reproduced in Fig.~\ref{pol_superlatt}. 
The two models globally reproduce the same trend, however the second one built on the relaxed pattern is better in agreement 
with the calculations.
In fact, from the parameters related to the first curve, with smaller curvature, a polar state is admitted for all $m$ instead of the condition $m\geq2$. 
On one hand, the quantitative inconsistencies between the two approaches, both in the distortion pattern and in ferroelectric properties of the superlattice,
point out that a simple model only based on pure bulk quantities can fail in reproducing the dynamics of the system because of 
strong coupling between the two different $B$-cations at play.
On the other hand, the perfect agreement obtained between the ``second'' model and first-principles calculations (Fig.~\ref{pol_superlatt}) confirms the validity of the hypothesis of uniform
$P$ in the superlattice, i.e. electrostatic energy cost equal to zero, due to the fact that the polar distortion of Zr preserves the charge transfer along Zr-O-Ti-O-Ti-O chains.
Moreover, $P$ asymptotically approaches the value of $P4mm$-BaTiO$_3$, that is of about $\simeq 34~\mu C/cm^{2}$ as calculated via DFT.
	
\begin{table}[b]
\centering
\footnotesize
\begin{tabular}{|ccc||cc|cc|}
\hline
\multicolumn{3}{|c||}{BaTiO$_3$} & \multicolumn{4}{c|}{BaZrO$_3$} \\
\hline
$\alpha_T$ & $\beta_T$ & $\gamma_T$ & $\alpha_Z^{(a)}$ & $\beta_Z^{(a)}$  & $\alpha_Z^{(b)}$ & $\beta_Z^{(b)}$ \\
-0.054 & 1.8x10$^{-5}$ & 3.4x10$^{-9}$ & 0.048 & 1.37x10$^{-5}$ & 0.073 & 2.26x10$^{-5}$  \\
\hline
\end{tabular}
\caption{\footnotesize Values of the parameters used in the $U_{BTO}(P)$ and $U_{BZO}(P)$
expansion as resulted from a fitting procedure. For BaZrO$_3$, we reports the parameters for both the constructions: 
(a) from the minimization procedure at fixed $P$ (blue curve in Fig.~\ref{E_BTO_BZO});
(b) from the empirical approach (green curve in Fig.~\ref{E_BTO_BZO}).} 
\label{fit_parameters}
\end{table}
 
These results, in particular the cooperative BaTiO$_3$-like atomic motion along the Zr-O-Ti-O-Ti-O chains, are the manifestation of the ``chain-like'' nature of the distortion. 
In fact, we already discussed in Sec.~\ref{btz_super} how the break of correlated Ti-O chains introduced by the polar inactive Zr prevents the propagation of the original BaTiO$_3$ distortion 
in Ba(Ti,Zr)O$_3$ solid solutions. Nevertheless, when a minimun number of such correlated chains is preserved, the 
substitution of one atom of Ti with one of Zr in those chains does not kill the polarization.
On the contrary, the electrostatic coupling forces the system to sustain a homogeneous $P$ by making Zr polar active. \\

However, previous calculations and physical interpretation have been based on superlattices with decreasing Zr-concentration. Therefore, in order to further prove the crucial role played by 
the local ``chain-like'' correlation over the increasing BTO-concentration, we have better distinguished the effects arising from the composition and from the imposed atomic ordering.
Thus we carried DFT calculations on 1x2x$\mathit{L}$ and 2x2x$\mathit{L}$
ordered supercells as we did for the 1x1x$\mathit{L}$ case. These two supercells, composed by alternating single chains of BZO/$m$BTO and BTO/$n$BZO,
allow to keep the total composition fixed to 50\% of Zr and Ti 
and to locally preserve the imposed ``chain-like'' atomic ordering, as reproduced in Fig.\ref{super_chain}(b,c).

For both the arrangements, it is not surprising to get no polar phases for $L=2$
from the inversion symmetry breaking along the $z$ direction. In fact, by periodicity, these structures correspond to the
columnar and rocksalt orders described in Sec.~\ref{btz50_solution}, Fig.~\ref{super_12}.
Relevant results arise from $L\geq3$, since a spontaneous polarization appears along the BZO/$m$BTO chains
driven by the (Zr,Ti)-motion against the oxygens inside the global BaTi$_{0.50}$Zr$_{0.50}$O$_3$ matrix.
As in the 1x1x$\mathit{L}$ superlattice, in fact, we described the atomic displacements as a cooperative motion around CM and
we found out that most of the polar distortion comes from the $B$-cations stacked on the local BZO/$m$BTO chains.
We also obtained no negligible $d_{33}$ piezoelectric coefficient, but smaller than in the 1x1x$\mathit{L}$ and pure $P4mm$-BaTiO$_3$ cases, as reported in Table~\ref{PdvsL} 
(because of the high computational cost for these supercells, we performed DFPT calculations only for the case $L$=3).
A simple explanation could be
that the coexistence of the BZO/$m$BTO and BTO/$n$BZO chains, that behave differently
since the cations in the latter one do not participate to the polarization, could globally prevent the variation of $P$. 
Noteworthy is that the polar $P4mm$ phase within the 2x2x$\mathit{L}$ structure is stable. \\

The main finding from the comparison of results coming from the three different structures is the polar activation of zirconium in
 Ba(Ti,Zr)O$_3$ system as a function of specific atomic ordering, that is local
BZO/$m$BTO chains with $m\geq2$, and independent of the total composition of the matrix.
Therefore, the analysis and physical interpretation of the dynamics contoducted through the previous electrostatic model remains valid also for the 1x2x$\mathit{L}$ and 2x2x$\mathit{L}$ cases: 
Zr atom is locally activated by a Zr-Ti coupling to avoid polarization gradients and electrostatic energy cost. 
But then, the gain of energy related to polarizing ferroelectric BaTiO$_3$ is decreased by the energy cost to polarize BaZrO$_3$.\\

From the $U(P)$ curves of the pure compounds in Fig.~\ref{E_BTO_BZO} and trends reported in Fig.~\ref{pol_superlatt}, 
the maximum Zr-concentration to get a ferroelectric system seems to be around 30\% in agreement with experimental observations. 
Beyond this concentration, the results are line with the observed relaxor behavior of BaTi$_{1-y}$Zr$_{y}$O$_3$ and 
existence of polar nanoregions associated to local fluctuation of $y$, that allows for correlated polar motions,
as described in Refs.[\cite{raman_BTZ_kreisel,BTZ_maiti,laulhe_1}]. 
Maiti \emph{et al.}, in their experimental investigation of the correlation between structure and property in Ba(Ti,Zr)O$_3$ ceramics\cite{BTZ_maiti, raman_BTZ_maiti}, point out that
 a gradual incorporation of Ti$^{4+}$ into the nonpolar BaZrO$_3$ lattice results in the evolution of relaxation behavior because of the increasing amount of 
ordering and density of nano-sized Ti$^{4+}$-rich polar regions in the Zr$^{4+}$-rich matrix. Moreover, C. Laulh\'e \emph{et al.} suggest that the observation of such 
polar nanoregions consist in 1D-chains with correlated Ti displacements \cite{laulhe_1}. 

Accordingly, by means of combined DFT calculations and phonomenological model, 
we have provided a complementary microscopic information. In particular, the behavior of the polar-clusters in a global paraelectric-matrix, like the case 
of the investigated BaTi$_{0.50}$Zr$_{0.50}$O$_3$ composition, can due to the cooperative motion of Zr and Ti atoms and not only by isolated Ti-dipoles.

\section{Discussion}
\label{discussion}

Properties of (Ba,Ca)TiO$_3$ and Ba(Ti,Zr)O$_3$, discussed in Sec.~\ref{bct_solid} and \ref{btz_solid}, can be better 
understood in relationship with those of the pure BaTiO$_3$, CaTiO$_3$, CaZrO$_3$ and BaZrO$_3$ parent compounds discussed in Sec.~\ref{parents_compound}. 
In fact, these two solid solutions result from the mixing of systems with very different dynamics. 
This turns out to be a crucial point regarding the suitableness of the \emph{ab-initio} method employed, VCA vs direct supercell calculations.

When going from BaTiO$_3$ to CaTiO$_3$ in (Ba,Ca)TiO$_3$, the main feature is the mixing of the so called $B$-driven and $A$-driven ferroelectricity associated to the two
parent compounds, respectively. 
Cubic BaTiO$_3$, in fact, exhibits strong and highly polar instabilities that expand over the entire $\Gamma$-$X$-$M$ planes of the Brillouin zone, 
as shown in Fig.~\ref{bulk_phon}(a). They are 
determined by the destabilizing Ti-O interaction resulting in the
relative motion of Ti atoms with respect to oxygens. Particularly, this polar distortion requires cooperative atomic displacements along Ti-O chains.
From the electronic point of view, such instability is ascribed to the hybridization between the O $2p$ and Ti $3d$ states\cite{cohen_BTO}. 
The sizaeble stiffness of Ba determines its negligible contribution to the distortion instead. 
Related values of the ``on-site'' and interatomic force constants are reported in Tables~\ref{sfc_bulk} and~\ref{ifc_bulk_local}.
These mechanisms stabilize the rhombohedral $R3m$-phase, as reported in Fig.~\ref{bulk_phases}.
On the contrary, CaTiO$_3$, in its cubic phase, hosts a strong polar instability confined at $\Gamma$ and mostly related to a polar motion of Ca and O. 
In fact, we found that a strong destabilizing interaction occurs between these two atoms (Table~\ref{ifc_bulk_local}). 
Such Ca driven instability is ascribed to steric effects as the smaller size and greater softness of Ca than Ba (Table~\ref{sfc_bulk}). 
In fact, same dynamical properties characterize also CaZrO$_3$ where the covalency between Zr and O is even further reduced\cite{cohen_BTO}. 
Moreover, the associated reduced volume of CaTiO$_3$ with respect to BaTiO$_3$ 
makes the Ti-O interaction mainly repulsive and Ti atoms much more stiffer. 
These mechanisms favor the tetragonal $P4mm$-phase over the $R3m$-one. However, the ground-state is determined by the strong 
AFD instabilities present along the $M-R$ line. See Figs.~\ref{bulk_phon}(b) and~\ref{bulk_phases}.
Accordingly, as obtained from supercell calculations for BaTiO$_3$-rich compositions, 
the mixing of the two $A$-cations in Ba$_{1-x}$Ca$_{x}$TiO$_3$ solid solutions does not make barium active in the polar distortion, but introduces a cooperative 
motion of Ca-atoms together with Ti-atoms against the oxygens-cage. 
In particular, a ferroelectric ground-state is predicted up to $x=0.50$.
Such prediction, has been achieved within both supercell-based and VCA approaches. 
Moreover, interesting trends in (Ba,Ca)TiO$_3$ are achievable within this approximation. 
In fact, as it is expected from the dynamical properties and energetics of the two parent compounds, the increasing Ca-composition favors 
 $P4mm$-like polar phases over the $R3m$ one.
In the Ba-rich region, 
this behaviour makes the three ferroelectric states very close in 
energy 
and gives rise to increasing piezoelectric response as the energy landscape in terms of polarization orientation becomes more and more isotropic (Figs.~\ref{bct_vca_energy} and~\ref{bct_vca_pol}). 
These properties actually occur for Ba$_{0.875}$Ca$_{0.125}$TiO$_3$, as obtained from the use of supercells. 
However, within VCA the distinction of the dynamics of Ba and Ca atoms is not accessible, because only a final average effect from the two cations at the $A$-site is reproduced, as 
shown in Fig.~\ref{acell_vca_bct}(c). 
This method forces the change of character of the mechanisms at play without allowing distinction of the opposite nature of Ba and Ca.
In fact, Ca is actually polar active for each $x$.
Such limitation is evident in the polarization trend, that is decreasing with Ca-concentration within VCA, while it is increasing 
within the supercells, and in the finding of the concentrantion where the inversion of polar phases takes place. 

When going from BaTiO$_3$ to BaZrO$_3$ in Ba(Ti,Zr)O$_3$, the scenario is completely different.
BaZrO$_3$, in fact, does not host any polar instability. Zr atom is quite stiff and the 
Zr-O interaction is strongly repulsive, as reported in Tables~\ref{sfc_bulk} and~\ref{ifc_bulk_local}. 
From the electronic point of view, this can be ascribed to the fact that Zr occupies $4d$ states less hybridized with the $2p$ states of O.
Therefore, at first, 
low Zr-doping further stabilizes the $R3m$-phase in BaTi$_{1-y}$Zr$_{y}$O$_3$ because of the larger ionic radius of Zr than Ti. However, 
the spontaneous polarization is slightly reduced.
Then, for higher Zr-concentration the system tends to globally reduce the polar distortion. In fact, the presence of Zr on the $B$-site introduces breaks along the correlated Ti-O-Ti-O chains,
by preventing the preservation of polarization. Such mechanisms are evident both for $y=0.125$ and $y=0.50$ supercells compositions.
For instance, in BaTi$_{0.50}$Zr$_{0.50}$O$_3$, the energetically favored  $Fm\bar3m$-rocksalt configuration is stable.
Nevertheless, ferroelectricity can be locally preserved in Ti-rich regions. 
Moreover, by building different supercell environments based on BaZrO$_3$/$m$BaTiO$_3$ chains, we found that 
as soon as Zr experiences at least 2 subsequent Ti atoms, it becomes active in the polarization and a cooperative motion of the two $B$-cations is observed. 
The reason of such behavior, explained via a basic electrostatic model, is that Zr is locally activated
in order to preserve homogeneous polarization along the Zr-O-Ti-O-Ti-O chain and to minimize the electrostatic energy cost. 
Such important role played by local structures on the Zr-Ti coupling and differences in the electronic configuration of the two $B$-cations
 make VCA not suitable for the Ba(Ti,Zr)O$_3$ system. In fact, predictions achieved via supercell-based calculations concerning both
 the dynamics and energetics are not reproduced by VCA. 

In both (Ba,Ca)TiO$_3$ and Ba(Ti,Zr)O$_3$, we found that 
the two types of cations introduced at the $A$ and $B$ sites, respectively, behave very differently and can therefore hardly be described by the same averaged virtual atom.
Although providing some trends, VCA is therefore not suitable for reproducing and understanding the microscopic physics of (Ba,Ca)(Ti,Zr)O$_3$ solid solutions. 
This clarifies why recent attempts \cite{nahas} to study BCTZ from a VCA-based effective Hamiltonian required to adjust by hands some of the parameters initially fitted from first-principles 
in order to reproduce experimental data.

\section{Conclusions}
\label{conclusion}

We have investigated the dynamical properties of chosen compositions of (Ba,Ca)(Ti,Zr)O$_3$ solid solutions and the four parent compounds by means of first-principles calculations. 

In (Ba,Ca)TiO$_3$ the mixing of $B$- and $A$-type ferroelectricity is the key role to get competitive 
polar phases. In fact, the two parent compounds have a reverse energy sequence of ferroelectric states and a crossover between them is required to achieve the inversion. 
This behaviour enhances the piezoelectric response. Morever, we have predicted the existence 
of a ferroelectric ground-state characterized by a cooperative (Ca,Ti)-motion at least up to the 50\% composition of the solid solution. 

Different behaviour charaterizes Ba(Ti,Zr)O$_3$, where the appearence of ferroelectricity is strongly dependent on the local
atomic arrangements and composition. Although the cubic phase is highly stable for BaZrO$_3$, by the use of a simple electrostatic model, we found out that the presence of  
(Zr,$m$Ti)-chains favors correlation of polar distortions in order to preserve an homogeneous polarization along the chain. 
As a consequence, even if a ferroelectric ground-state can be observed up to the critical concentration of 30\% of zirconium, 
polar nano-regions can be locally preserved for larger concentrations. 

From a methodological point of view, we have provided a direct comparison between the virtual crystal approximation (VCA) and direct supercell calculations. 
We have demonstrated that the specific microscopic physics of the (Ba,Ca)TiO$_3$ and Ba(Ti,Zr)O$_3$ 
solid solutions imposes severe limitations to the applicability of the virtual crystal approximation. 
By construction, this approximation fails at reproducing both specific local arrangements and the same-site independent 
motion of the active atoms involved in the ferroelectric instabilities. Also, it oversimplifies the actual electronic band structure of the doped systems. 
These are key features for the overall behavior of BaTiO$_3$-based solid solutions that need to be addressed by means direct supercell calculations.

We think that our findings establish a solid guideline for further investigations in understanding
 more deeply properties of the interesting (Ba,Ca)(Ti,Zr)O$_3$ solid solutions and in designing new lead-free piezoelectrics.

\section{Acknowledgments}

This work is supported by the European project EJD-FunMat 2015, program H2020-MSCA-ITN-	2014. 
Computational resources have been provided by the Consortium des \`Equipements de Calcul Intensif (CECI) in Belgium. 

D.A. would like to thank people in the PhyTheMa group in ULg for useful discussions and technical support.

\end{document}